\documentclass[aps,prd,twocolumn,superscriptaddress,showpacs,amsmath,amssymb,nofootinbib,eqsecnum, preprintnumbers]{revtex4-1}
\usepackage{graphicx}
\usepackage{epstopdf}
\usepackage{float}
\usepackage{color}
\usepackage[toc,page]{appendix}
\usepackage[usenames,dvipsnames]{xcolor}

\usepackage[utf8]{inputenc}

\newcommand{\be}{\begin{equation}}
\newcommand{\ee}{\end{equation}}
\newcommand{\ba}{\begin{eqnarray}}
\newcommand{\ea}{\end{eqnarray}}

\newcommand{\beq}{\begin{equation}}
\newcommand{\eeq}{\end{equation}}
\newcommand{\beqa}{\begin{eqnarray}}
\newcommand{\eeqa}{\end{eqnarray}}
\newcommand{\nn}{\nonumber}

\begin{document}

\title{Ultraspinning limits and super-entropic black holes}

\author{Robie A. Hennigar}
\email{rhennigar@uwaterloo.ca}
\affiliation{Department of Physics and Astronomy, University of Waterloo,
Waterloo, Ontario, Canada, N2L 3G1}

\author{David Kubiz\v n\'ak}
\email{dkubiznak@perimeterinstitute.ca}
\affiliation{Perimeter Institute, 31 Caroline St. N., Waterloo,
Ontario, N2L 2Y5, Canada}
\affiliation{Department of Physics and Astronomy, University of Waterloo,
Waterloo, Ontario, Canada, N2L 3G1}

\author{Robert B. Mann}
\email{rbmann@sciborg.uwaterloo.ca}
\affiliation{Department of Physics and Astronomy, University of Waterloo,
Waterloo, Ontario, Canada, N2L 3G1}

\author{Nathan Musoke}
\email{nmusoke@perimeterinstitute.ca}
\affiliation{Perimeter Institute, 31 Caroline St. N., Waterloo,
Ontario, N2L 2Y5, Canada}

%
\date{July 22, 2015}

\begin{abstract}
By employing the new ultraspinning limit we construct novel classes of black holes with
non-compact event horizons and finite horizon area and study their thermodynamics.
Our ultraspinning limit can be understood as a simple generating technique that consists of three steps: i) transforming the known rotating AdS black hole solution to a special coordinate system that rotates (in a given 2-plane) at infinity
ii) boosting this rotation to the speed of light iii) compactifying the corresponding azimuthal direction. In so doing we qualitatively change the structure of the spacetime since it is  no longer possible to return to a frame that does not rotate at infinity.
The obtained black holes
have non-compact horizons with topology of a sphere with two punctures. The entropy of some of these exceeds the maximal bound implied by the reverse isoperimetric inequality, such black holes are super-entropic.
\end{abstract}


\maketitle


\section{Introduction}

A fundamental result in the study of black holes is Hawking's theorem concerning the {\em topology} of black hole horizons~\cite{Hawking:1971vc}. Hawking showed that the two-dimensional event horizon cross sections of four-dimensional asymptotically flat stationary black holes satisfying the dominant energy condition necessarily have topology $S^2$.  This result indicates that asymptotically flat, stationary black holes in four dimensions are highly constrained systems.

More interesting black hole solutions are permitted in four and higher dimensions if one relaxes some of the assumptions going into Hawking's theorem. For example, since Hawking's argument relies on the Gauss--Bonnet theorem, it does not directly extend to higher dimensions. It is then not so surprising that higher-dimensional spacetimes permit a much richer variety of black hole topologies. The most famous example of this type is the black ring solution of Emparan and Reall which has horizon topology $S^2 \times S^1$ \cite{Emparan:2001wn}.  Despite the failure of Hawking's result in higher dimensions, Galloway and Schoen proved the less restrictive condition that the $(d-2)$-dimensional cross section of the event horizon (in the stationary case) and outer apparent horizons (in
the general case) are of positive Yamabe type, i.e., admit metrics of positive scalar curvature \cite{Galloway:2005mf}.

Another possibility is to relax asymptotic flatness.
 For example, in four-dimensional (locally) asymptotically  anti de Sitter (AdS) space the Einstein equations admit black hole solutions with the horizons being Riemann surfaces of any genus $g$~\cite{Vanzo:1997gw, Mann:1996gj, Lemos:1994xp, Cai:1996eg, Mann:1997iz}.  Higher-dimensional asymptotically AdS spacetimes are also known to yield interesting horizon topologies, for example, black rings with horizon topology $S^1 \times S^{d-3}$  \cite{Figueras:2014dta} and  rotating black hyperboloid membranes with horizon topology $\mathbb{H}^2 \times S^{d-4}$ \cite{Caldarelli:2008pz}.  More generally, event horizons which are Einstein manifolds of positive, zero, or negative curvature are possible in $d$-dimensional asymptotically AdS space \cite{Mann:1996gj,Birmingham:1998nr}.

Recently a new type of four-dimensional rotating black hole solution has been constructed in \cite{Gnecchi:2013mja} and elaborated upon in \cite{Klemm:2014rda,Hennigar:2014cfa} for both $\mathcal{N} =2$ gauged supergravity coupled to vector multiplets and Einstein--Maxwell-$\Lambda$ theory.  Supergravity solutions such as this are generically interesting since they correspond to string theory ground states, and therefore topics such as microscopic  degeneracy can be studied utilizing the AdS/CFT correspondence~\cite{Strominger:1996sh}.  Interest in this particular solution is further motivated by the fact that these black holes possess a non-compact event horizon of finite area (and therefore finite entropy), providing the first example of such
objects in the literature to date.  Topologically, the event horizon is a sphere with two punctures, demonstrating that the landscape of possible event horizon topologies is even richer than previously thought.

These black holes, in a sense, correspond to a new type of {\em ultraspinning limit} of the Kerr--Newman-AdS solution.
Ultraspinning black holes were first studied by Emparan and Myers  \cite{Emparan:2003sy} in an analysis focusing on the stability of Myers--Perry black holes \cite{Myers:1986un} in the limit of large angular momentum.  The analogous limit for rotating Kerr-AdS black holes is the case where the rotation parameter, $a$, approaches the AdS radius, $l$; however, the result of the limit is not unique and depends on how the limit is performed.  Caldarelli et al. \cite{Caldarelli:2008pz} considered the case where $a \to l$ keeping the physical mass $M$ fixed while simultaneously zooming in to the pole.  This limit is sensible only for $d \ge 6$ and yields a static black brane. Armas and Obers later showed that the same solution can be obtained by taking $a \to \infty$ while keeping the ratio $a/l$ fixed, their approach having the advantage of being directly applicable to dS solutions as well \cite{Armas:2010hz}.  Caldarelli et al. have also studied the $a\to l$ limit in the case of fixed $r_+$ while zooming into the pole  \cite{Caldarelli:2008pz, Caldarelli:2012cm}.  This prescription, valid for $d \ge 4$, yields a rotating black  hyperboloid membrane with horizon topology $\mathbb{H}^2 \times S^{d-4}$.  To avoid confusion in what follows we shall refer to the first mentioned AdS ultraspinning limit as the {\it{black brane limit}}, to the second as the {\it{hyperboloid membrane limit}}, and (for reasons that will become clear shortly) to the ultraspinning limit considered in this work as the {\it{super-entropic limit}}.  Interestingly, as shown in \cite{Klemm:2014rda}, the super-entropic limit coincides with the hyperboloid membrane limit near the poles, but globally they are distinct.

More recently, a simple technique was introduced in \cite{Hennigar:2014cfa} allowing one to perform the super-entropic limit directly from the Kerr--Newman-AdS solution.  The essence of this procedure is as follows:  one begins with the Kerr--Newman-AdS metric written in rotating-at-infinity coordinates then transforms the azimuthal coordinate $\phi$ and takes the $a \to l$ limit in a way that keeps the metric finite, but results in a non-compact azimuthal coordinate.  One can then compactify the new azimuthal coordinate to obtain the solution presented in \cite{Gnecchi:2013mja, Klemm:2014rda}.  The resulting metric rotates with the speed of light at infinity, and so the structure of the spacetime has been qualitatively changed through this limit.  This `{\em generating procedure}' can be applied to the singly rotating Kerr-AdS solution in $d$-dimensions \cite{Hennigar:2014cfa}, generalizing the original 4-dimensional solution \cite{Klemm:2014rda} to higher-dimensions.  The $d$-dimensional black holes resulting from this procedure have horizons that are topologically $(d-2)$-spheres with two punctures.

The analysis of the {\em extended phase space thermodynamics} (see, e.g., review \cite{Altamirano:2014tva}) for these unique black holes provided more motivation for their study.  Indeed, it was recently shown \cite{Hennigar:2014cfa} that in extended thermodynamic phase space, these black holes provide the first counterexample to the conjectured `{\em Reverse Isoperimetric Inequality}' \cite{Cvetic:2010jb}: the physical statement asserting that for a black hole of given thermodynamic volume the entropy will be maximal for the (charged) Schwarzschild AdS black hole.  As such, these black holes exceed their expected maximal entropy and so we refer to them as `{\em super-entropic}'.

The purpose of this paper is to further explore the applicability of the super-entropic limit. In particular, we shall investigate whether such a limit can be taken for multi-spinning black holes and/or combined with the traditional ultraspinning limits.
In so doing we will generate a broad class of new multi-spinning super-entropic black holes (with one super-entropic direction) in higher dimensions
and, in particular, obtain new super-entropic black holes in minimal gauged supergravity.
We find that while the black brane limit can be taken simultaneously in several directions (i.e. for several rotation parameters),
this seems impossible for the super-entropic limit.
While it seems that the black brane and super-entropic limits cannot be combined, we managed
to combine the super-entropic limit with the hyperboloid membrane limit, obtaining a new interesting solution that we describe in Appendix ~B.

Our paper is organized  as follows. We begin in Sec.~II with the discussion of singly-spinning super-entropic black holes: we
review how these solutions can be obtained by taking the super-entropic limit of singly-spinning Kerr-AdS black holes and discuss their
extended phase space thermodynamics. We then use the straightforward super-entropic limit procedure to obtain a broad class of new black hole solutions.  In particular, in Sec.~III we generate a new solution of minimal gauged supergravity, and in Sec.~IV generalize the singly spinning super-entropic black holes to the case of multiple rotations.
In all cases we discuss the extended phase space thermodynamics 
and investigate the isoperimetric ratio to determine whether the newly constructed black holes are super-entropic or not. Our conclusions are in Section V, after which we have three
appendices containing supplementary material about the various ultraspinning limits.  Appendix ~A is devoted to the black brane limit of multiply spinning Kerr-AdS black holes, Appendix ~B
to the hyperboloid membrane limit, and   Appendix ~C to the `uniqueness' of the `special rotating frame' employed  in the super-entropic limit procedure.

\section{Singly spinning super-entropic black holes}

\subsection{Super-entropic limit of Kerr-AdS black hole}

In what follows we shall construct new AdS black hole solutions by employing the novel super-entropic ultraspinning limit in which the rotation parameter $a$ attains its maximal value, equal to the AdS radius $l$. The procedure consists of the following steps. i) We start from a given rotating AdS black hole and, to eliminate any possible divergent terms in the metric that would prevent us from taking the $a\to l$ limit, recast it in a rotating-at-infinity coordinate system  that allows one to introduce  a rescaled azimuthal coordinate. ii) We then take the $a\to l$ limit, effectively `boosting' the asymptotic rotation to the speed of light. iii) Finally, we compactify the corresponding azimuthal direction. In so doing we qualitatively change the structure of the spacetime since it is  no longer possible to return to a frame that does not rotate at infinity. The obtained black holes have non-compact horizons with topology of a sphere with two punctures.
After analyzing some of their properties, we study the extended phase space thermodynamics of such black holes. 
As we shall see, they exceed the maximal entropy bound implied by the reverse isoperimetric inequality. Such black holes are  super-entropic.

Let us first demonstrate this procedure on the Kerr--Newman-AdS black hole in four dimensions~\cite{Carter:1968ks}.
We write the metric in the `standard Boyer--Lindquist form'~\cite{Hawking:1998kw},
\ba\label{KNADS2}
ds^2&=&-\frac{\Delta_a}{\Sigma_a}\left[dt-\frac{a\sin^2\!\theta}{\Xi}d\phi\right]^2
+\frac{\Sigma_a}{\Delta_a} dr^2+\frac{\Sigma_a}{S}d\theta^2\nonumber\\
&&+\frac{S\sin^2\!\theta}{\Sigma_a}\left[a dt-\frac{r^2+a^2}{\Xi}d\phi\right]^2\,,\nonumber\\
{\cal A}&=&-\frac{qr}{\Sigma_a}\left(dt-\frac{a\sin^2\!\theta}{\Xi}d\phi\right)\,,
\ea
where
\ba\label{KerrSigmaa}
\Sigma_a&=&r^2+a^2\cos^2\!\theta\,,\quad \Xi=1-\frac{a^2}{l^2}\,,
\quad S=1-\frac{a^2}{l^2}\cos^2\!\theta\,,\nonumber\\
\Delta_a&=&(r^2+a^2)\Bigl(1+\frac{r^2}{l^2}\Bigr)-2mr+q^2\,,
\ea
with the horizon $r_h$ defined by $\Delta_a(r_h)=0$. As written, the coordinate system rotates at infinity
with an angular velocity $\Omega_\infty =- a/l^2$  and
the azimuthal coordinate $\phi$ is a compact coordinate with range $0$ to $2 \pi$.
The choice of coordinates (\ref{KNADS2}), while convenient,
is not necessary to obtain the metric (\ref{KNADS2U}) below, as we demonstrate in Appendix ~\ref{AppC}.

We now want to take the limit $a\to l$.
To avoid a singular metric in this limit, we need only define a new azimuthal coordinate $\psi = \phi/\Xi$ (the metric is already written in coordinates that rotate  at infinity) and identify it with period $2\pi/\Xi$ to prevent a conical singularity.
After this coordinate transformation the $a\to l$ limit can be straightforwardly taken and we get the following solution:
\ba\label{KNADS2U}
ds^2&=&-\frac{\Delta}{\Sigma}\left[dt-l\sin^2\!\theta d\psi\right]^2
+\frac{\Sigma}{\Delta} dr^2+\frac{\Sigma}{\sin^2\!\theta}d\theta^2\nonumber\\
&&+\frac{\sin^4\!\theta}{\Sigma}\left[l dt-(r^2+l^2)d\psi\right]^2\,,\nonumber\\
{\cal A}&=&-\frac{qr}{\Sigma}\left(dt-l\sin^2\!\theta d\psi\right)\,,
\ea
where
\ba\label{1sdeltasig}
\Sigma=r^2+l^2\cos^2\!\theta\,,\quad \Delta=\Bigl(l+\frac{r^2}{l}\Bigr)^2-2mr+q^2\,.\quad
\ea
Note that coordinate $\psi$ is now a noncompact azimuthal coordinate, which we now choose to compactify by requiring that $\psi \sim \psi + \mu$.  The result is equivalent to the metric presented in \cite{Klemm:2014rda} for the case of vanishing magnetic and NUT charges, as can be seen directly using the following coordinate transformation:
\be
\tau=t\,,\quad p=l\cos\theta\,,\quad \sigma=-\psi/l\,,\quad L=\mu/l\,.
\ee
Originally, this solution was found as a limit of the Carter--Pleba\'nski solution and corresponds to the case where the angular quartic structure function has two
double roots  \cite{Gnecchi:2013mja, Klemm:2014rda}.

\subsection{Basic properties}

Although  the metric \eqref{KNADS2U} have been previously investigated insofar as its the basic properties \cite{Klemm:2014rda} and thermodynamics \cite{Hennigar:2014cfa} are concerned,  for completeness we review and elaborate upon them here.
We find that the metric \eqref{KNADS2U}  indeed describes a black hole, with horizon at $r=r_+$ (the largest root of $\Delta(r_+)=0$), and whose topology is that of a cylinder, i.e. a sphere with two punctures.  Indeed any fixed $(r,t)$ sections have the same topology: they are non-compact and  approach Lobachevsky space near the axis.  The $\theta=0,\pi$ axis is removed from the spacetime, and the
$\psi$ coordinate becomes null as $r\to\infty$.

We first note that there is a minimum value of the mass required for  horizons to exist.  Examining the roots of $\Delta$ in eq. \eqref{1sdeltasig} we find
\be
m \ge m_0 \equiv 2r_0\Bigl(\frac{r_0^2}{l^2}+1 \Bigr)\,,
\ee
where
\be
r_0^2 \equiv \frac{l^2}{3}\left[-1 + \left(4 + \frac{3 q^2}{l^2}\right)^{\frac{1}{2}} \right].
\ee
For $m > m_0$ horizons exist while for $m<m_0$ there is a naked singularity.  When $m=m_0$ the two roots of $\Delta$ coincide and the black hole is extremal.

Since
\be
g_{\psi \psi} = \frac{l^4 \sin^4{\theta}}{l^2\cos^2{\theta} + r^2}\left(2mr-q^2 \right)\,,
\ee
it follows (using $m>m_0$ and $r_+ > r_0$) that $g_{\psi\psi}$ is strictly positive outside the horizon, indicating that the spacetime is free of closed timelike curves.

To gain a deeper understanding of the spacetime, let us consider the geometry of constant $(t, r)$ surfaces.  The induced metric on such a surface reads,
\be
ds^2 = \frac{r^2 + l^2\cos^2\theta}{\sin^2\theta} d\theta^2 + \frac{ l^2\sin^4\theta(2mr - q^2)}{r^2+l^2\cos^2\theta}d\psi^2\,.
\ee
This metric appears to be ill-defined for $\theta=0,\pi$.  To ensure there is nothing pathological occurring near these points let us examine the metric in the small $\theta$ limit (due to symmetry, the $\theta = \pi$ limit will be identical).  We introduce the change of variables,
\be
\kappa = l(1-\cos\theta)\,,
\ee
and examine the metric for small $\kappa$.  This yields
\be
ds^2 = (r^2+l^2)\left[\frac{d\kappa^2}{4\kappa^2} +\frac{4(2mr-q^2) }{(r^2+l^2)^2} \kappa^2 d\psi^2\right]\,,
\ee
which is nothing but a metric of constant negative curvature on a quotient of the hyperbolic space $\mathbb{H}^2$.  This implies that the $t,r=const.$ slices are non-compact manifolds and that the space is free from pathologies near the poles.\footnote{The statement that these surfaces are non-compact should not be confused with the idea that they extend to $r=\infty$: they are, after all, a surface at $r=const.$.  The notion is better understood as meaning that there is infinite proper distance between any fixed $\theta \in (0,\pi)$ and either pole.}
In particular, {this} analysis applies to the case of the black hole horizon, for which,
\ba\label{sshorizonLimit1}
ds_h^2 = (r_+^2+l^2) \left[\frac{d\kappa^2}{4\kappa^2}+ \frac{4\kappa^2}{l^2} d\psi^2 \right]\,,
\ea
showing that the horizon is non-compact.

The above argument has allowed us to conclude that, near the poles, the spacetime is free of pathologies.  However, using this argument alone we cannot conclude anything definitive about what happens precisely at $\theta=0, \pi$.  Shortly we shall return to this question and move towards an answer through a study of geodesic motion in the spacetime.
The corresponding analysis indicates that the $\theta=0$ axis appears to be excised from the spacetime.

\begin{figure}[htp]
\includegraphics[width=0.8\linewidth, height=0.4\textheight]{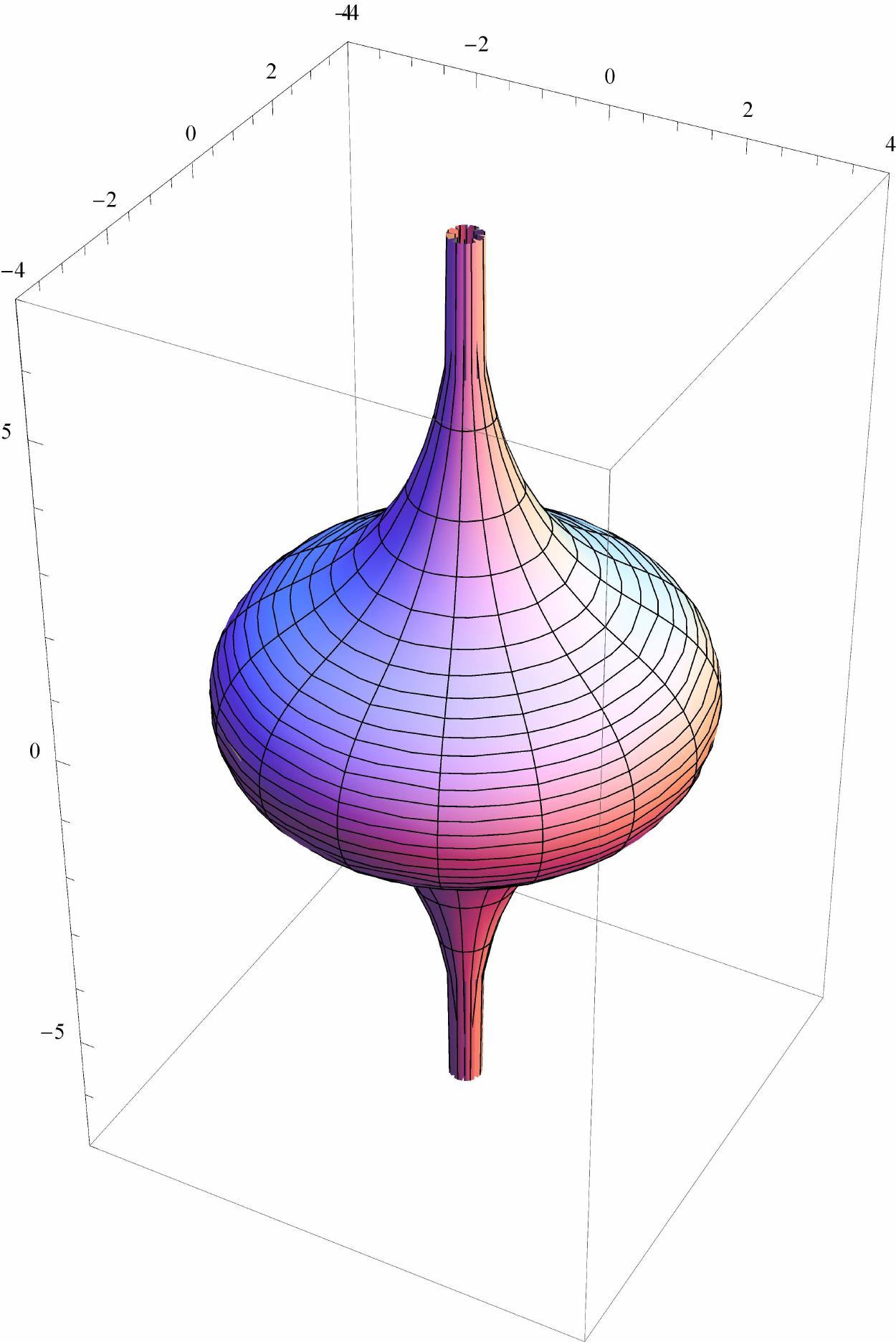}
\caption{{\bf Horizon embedding.} The horizon geometry of a 4d super-entropic black hole is embedded in $\mathbb{E}^3$ for the following choice of parameters: $q=0$, $l=1$, $r_+=\sqrt{10}$ and $\mu = 2\pi$.}\label{4dhorizon}
\end{figure}

To visualize the geometry of the horizon, we embed it in Euclidean 3-space \cite{Gnecchi:2013mja}.   The induced metric on the horizon is
\be
ds^2_h = g_{\psi\psi} d\psi^2 + g_{\theta\theta} d\theta^2\Big|_{r=r_+}\,.
\ee
We identify this line element with the line element in cylindrical coordinates,
\[ds_3^2 = dz^2+dR^2+R^2d\phi^2\,, \]
yielding
\ba
R^2(\theta) &=& \left(\frac{\mu}{2\pi }\right) g_{\psi\psi}\,,\label{Rofp}\\
\left(\frac{d z(\theta)}{d \theta} \right)^2 &=& g_{\theta\theta} - \left(\frac{d R(\theta)}{d\theta} \right)^2\,,\label{Rofp2}
\ea
where the prefactor in eq.~(\ref{Rofp}) comes from the manner in which we have compactified $\psi$.
Unfortunately, the resulting equations cannot be solved analytically. However it is straightforward to integrate them numerically for various values of $r_+, l$ and $q$,
as shown in Fig.~\ref{4dhorizon}.  We stress that the reader should not confuse the fact that $z(\theta)$ extends to $\pm \infty$ at the poles with the horizon extending to spatial infinity in the bulk spacetime.

The ergosphere is the region for which the  Killing vector $\partial_t$  is no longer timelike, given by
\be
\Delta-l^2\sin^4\!\theta \leq 0\,,
\ee
with equality corresponding to its outer boundary. Although at $\theta=0,\pi$ the ergosphere appears to touch the horizon, this does not take place since this axis is excised from the
spacetime as we shall see.

On the conformal boundary the metric \eqref{KNADS2U} takes the following form (the conformal factor being given by $l^2/r^2$)
\be
ds_{\text{bdry}}^2 = -dt^2 -  2l\sin^2\!{\theta} dt d\psi + \frac{l^2}{\sin^2\!{\theta}} d\theta^2
\ee
and we see that $\psi$ becomes a null coordinate  there.
Writing again $\kappa = l(1-\cos\theta)$,  the small $\kappa$ limit gives
\begin{equation}
ds_{\text{bdry}}^2 = -dt^2 - 4l\kappa d\psi dt + \frac{l^2}{4\kappa^2}d\kappa^2\,,
\end{equation}
which is nothing else but an AdS$_3$ written as a Hopf-like fibration over $\mathbb{H}^2$.
Due to the symmetry of the metric, an identical result holds for $\theta=\pi$.
This shows that there is no pathology near the poles while the poles themselves are excised from the boundary (see next subsection).

In fact, more generally for any fixed $r>r_+$, and after a substitution $\sin\theta =e^{-y}$, the expansion for small $\theta$ i.e. large $y$ gives
in the leading order
\be
  ds^2 = -\frac{\Delta}{r^2+l^2} dt^2 + (r^2+l^2)dy^2 + \frac{2l\Delta e^{-2y}}{r^2+l^2} dt d\psi+\dots\,.
\ee
To leading order in $e^{-2y}$, this metric is AdS$_3$ with $\psi$ a null coordinate, indicating that as we approach the poles the coordinate $\psi$ becomes null
(the component $g_{\psi\psi}$ vanishing as $e^{-4y}$). Retention of this latter term yields
(\ref{sshorizonLimit1}) as $r \to r_+$.

\subsection{Geodesics and the symmetry axis}\label{geodesics}

In order to understand the role of the symmetry axis $\theta=0,\pi$, we shall now study the geodesics.
The geometry admits a closed conformal Killing--Yano 2-form, $h=db$,
\be
b=(l^2\cos^2\!\theta-r^2)dt-l(l^2\cos^2\!\theta-r^2\sin^2\!\theta)d\psi\,,
\ee
inherited from the Kerr-AdS spacetime. Such an object guarantees separability of the Hamilton--Jacobi, Klein--Gordon, and Dirac equations in this background. In particular,
it generates a Killing tensor $k_{ab}=(*h)_{ac}(*h)^{c}{}_b$, $\nabla_{(a}k_{bc)}=0$, whose existence implies a Carter  constant of motion \cite{Carter:1968rr}, $k_{ab}u^au^b$, rendering geodesic motion (with 4-velocity $u^a$) completely integrable.

The fastest way to obtain the explicit expressions for the 4-velocity is to separate the Hamilton--Jacobi equation \cite{Carter:1968rr}
\be\label{HJ}
\frac{\partial S}{\partial \lambda}+g^{ab}\frac{\partial S}{\partial x^a}\frac{\partial S}{\partial x^b}=0\,,
\ee
where the inverse metric to \eqref{KNADS2U} reads
\ba
\partial_s^2&=&-\frac{1}{\Sigma\Delta}\left[(r^2+l^2)\partial_t+l\partial_\psi\right]^2
+\frac{\Delta}{\Sigma} \partial_r^2+\frac{\sin^2\!\theta}{\Sigma}\partial_\theta^2\nonumber\\
&&+\frac{1}{\Sigma\sin^4\!\theta}\left[l\sin^2\!\theta\partial_t+\partial_\psi\right]^2\,
\ea
and where one can identify $\partial S$ with the momentum 1-form $u$
\be\label{id}
\partial_aS=u_a\,.
\ee

We seek an additive separated solution (with the constants ${\cal E}, h, \sigma=-u^2$ corresponding to explicit symmetries)
\be\label{S}
S=\sigma \lambda-{\cal E} t+h\psi +R(r)+\Lambda(\theta)\,,
\ee
giving from (\ref{HJ})
\ba
&&\sigma-\frac{1}{\Sigma\Delta}\left[-(r^2+l^2){\cal E}+lh\right]^2
+\frac{\Delta}{\Sigma} R'^2+\frac{\sin^2\!\theta}{\Sigma}\Lambda'^2\nonumber\\
&&+\frac{1}{\Sigma\sin^4\!\theta}\left[h-l\sin^2\!\theta {\cal E}\right]^2=0\,,\quad
\ea
where $R'=dR/dr$ and $\Lambda'=d\Lambda/d\theta$. Multiplying by $\Sigma$ and
reshuffling the terms, we obtain
\ba\label{sep}
C&=&-\sigma r^2+
\frac{1}{\Delta}\left[-(r^2+l^2){\cal E}+lh\right]^2
-\Delta R'^2\nonumber\\
&=&\sin^2\!\theta\Lambda'^2+\sigma l^2\cos^2\!\theta+\frac{1}{\sin^4\!\theta}\left[h-l\sin^2\!\theta {\cal E}\right]^2\!,\qquad
\ea
where $C$ is   {\em Carter's constant}, the additional (hidden) integral of geodesic motion.

Hence the geodesic 4-velocity ($u_t=-{\cal E}, u_\psi=h$) is given by
\ba
\dot t&=&\frac{{\cal E}(2mr-q^2)l^2}{\Sigma \Delta}+\frac{lh\bigl(\Delta-\sin^2\!\theta(r^2+l^2)\bigr)}{\Sigma\Delta\sin^2\!\theta}\,,\nonumber\\
\dot \psi&=&\frac{h\bigl(\Delta-\sin^4\!\theta l^2\bigr)}{\Sigma \Delta \sin^4\!\theta}-\frac{l{\cal E}\bigl(\Delta-\sin^2\!\theta(r^2+l^2)\bigr)}{\Sigma\Delta\sin^2\!\theta}
\,,\nonumber\\
\dot r&=&\frac{\sigma_r}{\Sigma}\sqrt{\left[lh -(r^2+l^2){\cal E}\right]^2 {-\Delta C}-\sigma \Delta r^2}\,,
\label{geoeqs}\\
\dot \theta&=&\frac{\sigma_\theta\sin\theta}{\Sigma}\sqrt{ {C} -\frac{1}{\sin^4\!\theta}\left[h-l\sin^2\!\theta {\cal E}\right]^2-\sigma l^2\cos^2\!\theta}\,,\nonumber
\ea
where $\sigma_r=\pm$ and $\sigma_\theta=\pm$ are independent signs.

To fully understand these geodesics a further analysis going beyond the scope of this paper is required (as in \cite{Hackmann:2010zz}). In what follows we limit ourselves to presenting an argument showing that the symmetry axis $\theta=0,\pi$ cannot be reached by null geodesics ($\sigma=0$) emanating from the bulk in a finite affine parameter. This indicates that
the axis is some kind of a `boundary' that is to be excised from the spacetime.

Let us probe  the behavior close to $\theta=0$ (the discussion for $\theta=\pi$ is due to the symmetry analogous).
Consider `ingoing' null geodesics  for which $\theta$ decreases. For any finite value of $C$,
it is obvious from the expression underneath the square root in the last equation \eqref{geoeqs} that when $h\neq 0$, $\theta=0$ cannot be reached (the term $[h-l{\cal E}\sin^2\!\theta]^2/\sin^4\!\theta$ dominates for small $\theta$ driving the square root imaginary).

Consider next $h=0$. Then we have
\be\label{hdot}
\dot \theta=-\frac{\sin\theta}{\Sigma}\sqrt{C-l^2{\cal E}^2}\,.
\ee
It is straightforward to show from the third equation in (\ref{geoeqs}) that
there exists a constant $C=C_*>0$ and $r=r_*>r_+$ such that $\dot r(r_*)=0$; or in other words there exists a constant-$r$ surface along which such photons are confined. Such geodesics will spiral towards
$\theta=0$  with $\dot \psi\neq 0$. For small $\theta$ we obtain   $\dot \theta/\theta\approx -b^2=-\sqrt{C_*-l^2{\cal E}^2}/r_*^2=$constant, i.e., $\theta \to e^{-b^2\tau}$.  Photons moving on constant $r=r_*$ surfaces spiral toward  $\theta=0$ in infinite affine parameter.
Moreover, using the first equation \eqref{geoeqs} together with \eqref{hdot}, we have
\be
\frac{d\theta}{dt}=-k\sin\theta\,,\quad k=\frac{\Delta_*\sqrt{C_*-l^2{\cal E}^2}}{{\cal E}l^2(2mr_*-q^2)}>0\,.
\ee
Hence, starting from some finite $\theta_0$, we have
\be
t=-\frac{1}{k}\int\frac{d\theta}{\sin\theta}=-\frac{1}{k}\ln\Bigl(\tan\frac{\theta}{2}\Bigr)+\mbox{const.}
\ee
Evidently, as $\theta$ approaches zero, $t\propto -\frac{1}{k}\ln\theta\to \infty$; the axis is reached in infinite coordinate time $t$.
Hence photons of this type can never reach the symmetry axis.\footnote{For comparison, let us review here the behavior of radial geodesics in AdS space. Writing the metric in static coordinates, $ds^2=-fdt^2+dr^2/f$, $f=1+r^2/l^2$, we have 2 constants of motion $u^2=-\sigma$ and $u_t=-\epsilon$, giving
\be
\dot t=\frac{\epsilon}{f}\,,\quad 
\dot r=\pm\sqrt{\epsilon^2-\sigma f}\,.
\ee
Specifically, radial null geodesics ($\sigma=0$) starting from $r=0$ reach AdS boundary situated at $r=\infty$ in infinite affine parameter, $\tau=r/\epsilon\to \infty$, but (integrating $dr/dt=f$) at finite coordinate time $t=l\arctan(r/l)=\pi l/2$.}

The final possibility is that (while $h=0$) the coordinate $r$ changes as the photon approaches $\theta=0$.
Dividing the last two equation in \eqref{geoeqs} and introducing the following dimensionless quantities:
\be
x=\frac{r}{l}\,,\quad A=\frac{2m}{l(1-l^2{\cal E}^2/C)}>0\,,\quad B=\frac{q^2}{2ml}\,,
\ee
we find that
\ba\label{fatal1}
\int\frac{d\theta}{\sin\theta}=\ln\left(\tan\frac{\theta}{2}\right)
=-\sigma_r\int\frac{dx}{\sqrt{P(x)}}\,.\qquad
\ea
where $P(x)$ is the fourth-order polynomial given by
\be
P(x)=A(x-B)-(1+x^2)^2\,.
\ee
It is easy to see that $P(x)$ can have at most 2 positive roots $0<x_1<x_2$ and that geodesic motion occurs for  $r=xl$ obeying
$x_1\leq x\leq x_2$. The case $x_1=x_2$ corresponds to motion on fixed $r=r_*$ discussed in the previous paragraph.
To reach $\theta=0$, the l.h.s. of Eq.~\eqref{fatal1} diverges as $\ln\theta$. However, in the region of allowed motion, the
r.h.s. of \eqref{fatal1} remains finite (as only simple roots of $P(x)$ occur). This excludes the final possibility that
the axis $\theta=0$ can be reached by null geodesics emanating from some finite $\theta_0$ in the bulk.

 Finally, a much simpler argument, based on studying null geodesics on the conformal boundary, indicates that the axis of symmetry is in fact removed from the spacetime. Writing $\sin\theta=e^{-y}$, the metric on the conformal boundary reads
\be
ds^2=-dt^2+l^2dy^2+2le^{-2y}dtd\psi\,.
\ee
The geodesic motion on this space admits 3 constants of motion $u^2=-\sigma$, $u_t=-{\cal E}$ and $u_\psi=h$, giving the following 3 equations for null geodesics:
\ba
\dot t&=&\frac{h}{l}e^{2y}\,,
\quad
\dot \psi=\frac{e^{4y}}{l^2}(h-{\cal E}le^{-2y})\,,
\nonumber\\
\dot y&=&\pm \frac{e^{2y}}{l^2}\sqrt{h(2{\cal E}le^{-2y}-h)}\,.
\ea
From the last equation it is obvious that no null geodesic emanating from finite $y_0$ can reach the pole $y=\infty\ (\theta=0)$ on the conformal boundary.

To summarize, the above arguments clearly demonstrate that the symmetry axis $\theta=0,\pi$ is actually not part of the spacetime and represents instead some kind of a boundary.
It is an interesting question as to whether such a boundary has similar properties to those of the  boundary of AdS space (cf. footnote 2).

\subsection{Thermodynamics and the Reverse Isoperimetric Inequality}

We shall now study the thermodynamics of the obtained ultraspinning black hole \eqref{KNADS2U}. We do this in an {\em extended phase space} framework  \cite{Altamirano:2014tva},  where the cosmological constant is identified with the thermodynamic pressure according to
\be\label{eq:pressure}
P = -\frac{1}{8\pi}\Lambda = \frac{(d-1)(d-2)}{16\pi l^2}\,,
\ee
in $d$ spacetime dimensions, with its conjugate quantity  treated as thermodynamic volume $V$.
The first law of black hole thermodynamics then reads
\be\label{eq:first-law}
\delta M = T\delta S +\sum_i \Omega_i\delta J_i + \Phi\delta Q + V \delta P\,,
\ee
a result supported by geometric arguments \cite{Kastor:2009wy}. Note that the mass of the black hole $M$ is no longer interpreted as internal energy but rather as chemical enthalpy \cite{Kastor:2009wy}.
The angular velocities $\Omega_i$  and the electric potential $\Phi$ are measured with respect to infinity.
The corresponding Smarr relation
\be\label{Smarr}
\frac{d-3}{d-2}M=TS+\sum_i \Omega_i J_i+\frac{d-3}{d-2}\Phi Q-\frac{2}{d-2}VP\,,
\ee
can be derived from a scaling (dimensional) argument~\cite{Kastor:2009wy}.

A remarkable property of the thermodynamic volume is that (prior to the cases studied in~\cite{Hennigar:2014cfa}) for all black holes studied to date it satisfies what is known as the {\em reverse isoperimetric inequality} \cite{Cvetic:2010jb}.  Indeed, it was conjectured in~\cite{Cvetic:2010jb}
that the isoperimetric ratio
\be\label{eq:ipe-ratio}
\mathcal{R}=\left(\frac{(d-1) {V}}{\omega_{d-2}}\right)^{\frac{1}{d-1}}\left(\frac{\omega_{d-2}}{ {A}}\right)^{\frac{1}{d-2}}
\ee
always satisfies $\mathcal{R} \ge 1$. Here ${V}$ is the thermodynamic volume, ${A}$ is the horizon area, and $\omega_d$ stands for the area of the space orthogonal to constant $(t,r)$ surfaces; in the
$d$-dimensional super-entropic spacetime it is
\be\label{omega}
\omega_{d} = \frac{\mu \pi ^{\frac{d-1}{2}}}{\Gamma\left(\frac{d+1}{2}\right)}\,,
\ee
due to the compactification of the `super-entropic azimuthal coordinate'; the result for a standard unit sphere is recovered upon setting $\mu=2\pi$.
This inequality deepens our mathematical understanding of black hole thermodynamics insofar as it places a constraint on the entropy of an AdS black hole.  Physically, this inequality is the statement that for a black hole of a given thermodynamic volume, the entropy will be maximal for the (charged) Schwarzschild-AdS black hole.

In the framework of extended phase space thermodynamics the thermodynamic quantities associated with the solution \eqref{KNADS2U}
read  \cite{Klemm:2014rda,Hennigar:2014cfa}
\begin{align}
M&=\frac{\mu m}{2\pi}\,,\quad J=Ml\,,\quad
\Omega = \frac{l}{r_+^2+l^2}\,,\quad A=2\mu(l^2+r_+^2),  \nonumber
\\
S&= \frac{A}{4}, \quad T=\frac{1}{4\pi r_+}\left(3\frac{r_+^2}{l^2}-1-\frac{q^2}{l^2+r_+^2} \right), \nonumber \\
V &= \frac{r_+ A}{3},\quad \Phi=\frac{qr_+}{r_+^2+l^2}, \;\;\; Q=\frac{\mu q}{2\pi}\, .
\label{eq:thermo_properties}
\end{align}
Note that, due to the singular nature of the ultraspinning limit performed, these cannot be obtained by taking the $a \to l$ limit of the Kerr--Newman-AdS thermodynamic quantities. 

The isoperimetric ratio now reads (note that, due to the compatification  of $\psi$, the volume of the $2$-dimensional unit `sphere' in this spacetime is $2\mu$)
\be
\mathcal{R}=\left(\frac{r_+A}{2\mu}\right)^{1/3}\left(\frac{2\mu}{A}\right)^{1/2}
= \left(\frac{r_+^2}{r_+^2 + l^2}\right)^{1/6}< 1\,.
\ee
In other words, the obtained black holes provide a counterexample to the conjectured Reverse Isoperimetric Inequality---for a given thermodynamic volume their entropy exceeds that of the Schwarzschild-AdS black hole. As such, these black holes are super-entropic \cite{Hennigar:2014cfa}.

\subsection{Singly spinning super-entropic black holes in all dimensions}

To generalize the super-entropic black hole solution to higher dimensions, we start from the singly spinning $d$-dimensional Kerr-AdS geometry \cite{Hawking:1998kw},
\ba
ds^2 &=& -\frac{\Delta_a}{\rho^2_a}\left[dt - \frac{a}{\Xi} \sin^2\theta d\phi \right]^2 + \frac{\rho^2_a}{\Delta_a}dr^2 + \frac{\rho^2_a}{\Sigma_a}d\theta^2\qquad  \\
&+& \frac{\Sigma_a \sin^2\theta }{\rho^2} \left[a dt - \frac{r^2+a^2}{\Xi} d\phi \right]^2 + r^2 \cos^2\theta d\Omega_{d-4}^2\,, \nn
\ea
where
\ba
\Delta_a &=& (r^2+a^2)(1+\frac{r^2}{l^2}) - 2mr^{5-d}, \quad \Sigma_a = 1-\frac{a^2}{l^2}\cos^2\theta\,,\nn\\
\Xi &=& 1-\frac{a^2}{l^2}, \quad \rho^2_a = r^2+a^2\cos^2\theta\,.
\ea
 Replacing  $\phi = \psi \Xi$ everywhere and then taking the limit $a \to l$ we obtain
\ba\label{Singlyspinning}
ds^2&=&-\frac{\Delta}{\rho^2}(dt-{l}\sin^2\!\theta d\psi)^2+
\frac{\rho^2}{\Delta}dr^2+\frac{\rho^2}{\sin^2\!\theta}d\theta^2\quad \nonumber\\
&+&
\frac{\sin^4\!\theta}{\rho^2}[ldt-({r^2\!+\!l^2})d\psi]^2\!+\!r^2\cos^2\!\theta d\Omega_{d-4}^2\,,\qquad\ \
\ea
where
\ba
\Delta&=&\Bigl(l+\frac{r^2}{l}\Bigr)^2-2mr^{5-d}\,,\ \rho^2=r^2+l^2\cos^2\!\theta\,,\quad
\ea
and $d\Omega_{d}^2$ denotes the metric element on a $d$-dimensional sphere.  As before,  $\psi$ is a noncompact coordinate, which we now compactify via $\psi \sim \psi + \mu$.  It is straightforward to show that the metric
\eqref{Singlyspinning} satisfies the Einstein-AdS equations.  Horizons exist in any dimension $d> 5$ provided $m>0$ and in $d=5$ provided $m>l^2/2$.
We pause to remark that a method similar to that of~\cite{Gnecchi:2013mja,Klemm:2014rda} could be used
to  generate these solutions beginning with a $d$-dimensional generalization of   a Carter-Plebanski-like solution 
\cite{Klemm:1998kd} and then choosing its parameters so that the metric function has two double roots.  We do not explore this alternative here.

Similar to the 4-dimensional case, the solution inherits a closed conformal Killing--Yano 2-form from the Kerr-AdS geometry,
$h=db$, where
\be
b=(l^2\cos^2\!\theta-r^2)dt-l(l^2\cos^2\!\theta-r^2\sin^2\!\theta)d\psi\,.
\ee
This object guarantees complete integrability of geodesic motion as well as separability of the Hamilton--Jacobi, Klein--Gordon, and Dirac equations in this background;
see \cite{Frolov:2008jr} for analogous results in the Kerr-AdS case.
In particular, the geodesics can be discussed in a way analogous to the previous subsection.

Also the arguments concerning the behavior near the symmetry axis at $\theta = 0, \pi$ for the 4-dimensional case can be repeated here.   The induced metric on the horizon is
\ba\label{1shorizonMetric}
ds_h^2 &=& \frac{r_+^2 + l^2\cos^2\theta}{\sin^2\theta}d\theta^2 + \frac{\sin^4\theta(r_+^2+l^2)^2}{l^2 \cos^2\theta + r_+^2} d\psi^2
\nn\\
 &+& r_+^2 \cos^2\theta d\Omega_{d-4}^2\,,
\ea
and  introducing as before
$\kappa = l(1-\cos\theta)$
we find
\be\label{sshorizonLimit}
ds_h^2 = (r_+^2+l^2) \left[\frac{d\kappa^2}{4\kappa^2}+ \frac{4\kappa^2}{l^2} d\psi^2 \right] +r_+^2 d\Omega_{d-4}^2\,.\quad
\ee
This is a product  metric  of two spaces  $\mathbb{H}^2 \times S^{d-4}$  of constant  curvature; the horizons of these black holes are non-compact and have finite horizon area. Similar to the four-dimensional case, they have topology of a cylinder as the actual axis is excised from the spacetime.

The thermodynamic quantities for these black holes in extended phase space are given by,
\begin{eqnarray}\label{singlespin}
M&=&\frac{\omega_{d-2}}{8\pi} \left(d-2 \right)m  \,, \quad
J=\frac{2}{d-2}Ml\,,\quad \Omega=\frac{l}{r_+^2+l^2}\,,\nonumber\\
T&=&\frac{1}{4\pi r_+ l^2}\Bigr[ (d-5)l^2 + r_+^2(d-1)\Bigr]\,,\nonumber\\
S&=&\frac{\omega_{d-2}}{4}(l^2+r_+^2) r_+^{d-4}=\frac{A}{4}\,,\quad
V=\frac{r_+A}{d-1}\,,
\end{eqnarray}
with $\omega_d$ given by \eqref{omega}.
Here $\Omega$ is the angular velocity of the horizon and $J$ and $M$ have been computed via the method of conformal completion as the conserved quantities associated with the $\partial_\psi$ and $\partial_t$ Killing vectors, respectively. These quantities satisfy both the first law \eqref{eq:first-law} and the Smarr relation \eqref{Smarr} \cite{Hennigar:2014cfa}.

The isoperimetric ratio for these black holes reads
\ba
\mathcal{R} &=& \left(\frac{r_+ A}{\omega_{d-2}}\right)^{\frac{1}{d-1}}\left(\frac{\omega_{d-2}}{A}\right)^{\frac{1}{d-2}} \nn\\
&=& \left(\frac{r_+^2}{l^2+ r_+^2} \right)^{\frac{1}{(d-1)(d-2)}} < 1\,,
\ea
and so, similar to their 4-dimensional cousins, these black holes are also super-entropic.

\section{Black holes of minimal gauged supergravity}\label{SecSUGRA}

\subsection{Super-entropic limit}

Let us consider the general rotating charged black hole in five dimensions, a solution of minimal gauged supergravity constructed in \cite{Chong:2005hr},
\ba\label{SUGRAmetric}
ds^2&=& d\gamma^2 - \frac{2q \nu \omega}{\Sigma}+\frac{f \omega^2}{\Sigma^2} + \frac{\Sigma dr^2}{\Delta}
+ \frac{\Sigma d\theta^2}{S}\,,\quad
\nn\\
A &=& \frac{\sqrt{3}q\omega}{\Sigma}\,,
\ea
where we have defined
\ba
d\gamma^2&=&-\frac{S\rho^2dt^2}{\Xi_a\Xi_bl^2} + \frac{r^2\!+\!a^2}{\Xi_a} \sin^2\!\theta d\phi^2+ \frac{r^2\!+\!b^2}{\Xi_b}\cos^2\!\theta d\psi^2\,,\nn\\
\nu &=& b\sin^2\theta d\phi + a \cos^2\theta d\psi\,, \nn\\
\omega &=& \frac{S dt}{\Xi_a\Xi_b} -a\sin^2\!\theta\frac{d\phi}{\Xi_a} - b\cos^2\!\theta \frac{d\psi}{\Xi_b}\,,
\ea
and
\ba
S&=& \Xi_a\cos^2\theta +\Xi_b\sin^2\theta\,,
\nn\\
\Delta &=& \frac{(r^2+a^2)(r^2+b^2)\rho^2/l^2+q^2+2abq}{r^2}-2m\,,
\nn\\
\Sigma &=& r^2+a^2\cos^2\theta + b^2\sin^2\theta\,,\quad \rho^2=r^2+l^2\,,
\nn\\
\Xi_a &=& 1-\frac{a^2}{l^2}, \quad \Xi_b = 1-\frac{b^2}{l^2}\,,
\nn\\
f &=& 2m\Sigma -q^2+\frac{2abq}{l^2}\Sigma\,.
\ea
The black hole rotates in two different directions, corresponding to rotation parameters $a$ and $b$, parameter $q$ is related to the black hole charge.

Our goal here is to perform the super-entropic limit.  As we will see, this is only possible along one azimuthal direction, which we take to be the $\phi$-direction.
In so doing, we cannot apply directly the procedure used for the singly spinning solution to the metric in \cite{Chong:2005hr} since this metric is written in coordinates which do not rotate at infinity.  For this reason we perform the following coordinate transformation of $\phi$ and/or $\psi$:
\be\label{azirot}
\phi=\phi_R+\frac{a}{l^2}t\,,\quad \psi=\psi_R+\frac{b}{l^2}t\,,
\ee
where $\phi_R$ and $\psi_R$ are new `rotating at infinity coordinates'. We then have
\ba\label{nuomega2}
\nu&=&\frac{ab\sin^2\!\theta}{l^2}dt+b\sin^2\!\theta d\phi_R+a\cos^2\!\theta d\psi\nn\\
&=&\frac{ab}{l^2}dt+b\sin^2\!\theta d\phi_R+a\cos^2\!\theta d\psi_R\,,\\
\omega&=&\Bigl( 1-\frac{b^2}{l^2}\sin^2\theta\Bigr) \frac{dt}{\Xi_b}-\frac{a\sin^2\!\theta d\phi_R}{\Xi_a}-\frac{b\cos^2\!\theta d\psi}{\Xi_b}\qquad \nn\\
&=&dt-\frac{a\sin^2\!\theta d\phi_R}{\Xi_a}-\frac{b\cos^2\!\theta d\psi_R}{\Xi_b}\,.
\ea
At the same time, we find
\ba\label{gamma4}
d\gamma^2&=&
\frac{\sin^2\!\theta}{\Xi_a}\Bigl[(r^2+a^2)d\phi^2_R+\frac{2a}{l^2}(r^2+a^2)dtd\phi_R\nonumber\\
&&\ \ \ \ -\frac{dt^2}{l^2}\bigl(\rho^2-(r^2+a^2)\frac{a^2}{l^2}\bigr)\Bigr]\nonumber\\
&&+\frac{\cos^2\!\theta}{\Xi_b}\Bigl[(r^2+b^2)d\psi^2-\frac{\rho^2dt^2}{l^2}\Bigr]\,,
\ea
or,
\ba\label{gamm5}
d\gamma^2&=&
\frac{\sin^2\!\theta}{\Xi_a}\Bigl[(r^2+a^2)d\phi^2_R+\frac{2a}{l^2}(r^2+a^2)dtd\phi_R\nonumber\\
&&\ \ \ \ -\frac{dt^2}{l^2}\bigl(\rho^2-(r^2+a^2)\frac{a^2}{l^2}\bigr)\Bigr]\nonumber\\
&&+\frac{\cos^2\!\theta}{\Xi_b}\Bigl[(r^2+b^2)d\psi^2_R+\frac{2b}{l^2}(r^2+b^2)dtd\psi_R\nonumber\\
&&\ \ \ \ -\frac{dt^2}{l^2}\bigl(\rho^2-(r^2+b^2)\frac{b^2}{l^2}\bigr)\Bigr]\,,
\ea
giving the solution in transformed coordinates.

We are now ready to perform the super-entropic limit in the $\phi$-direction.  We begin by setting
\be\label{azirescale}
\varphi=\phi_R/\Xi_a\,,
\ee
while we keep $b$ as is, and then take the limit $a\to l$. We have $S \to \sin^2\!\theta \Xi_b$,
\ba\label{ho1}
\nu\to\nu_s&=&\frac{b}{l}\sin^2\!\theta dt+l\cos^2\!\theta d\psi\nn\\
&=&\frac{b}{l}dt+l\cos^2\!\theta d\psi_R\,,\\
\omega\to \omega_s&=&\Bigl( 1-\frac{b^2}{l^2}\sin^2\theta\Bigr) \frac{dt}{\Xi_b}-l\sin^2\!\theta d\varphi-\frac{b\cos^2\!\theta d\psi}{\Xi_b}
\qquad \nn\\
&=&dt-l\sin^2\!\theta d\varphi-\frac{b\cos^2\!\theta d\psi_R}{\Xi_b}\,,
\ea
and $d\gamma^2\to d\gamma_s^2$\,, where
\ba\label{ho2}
d\gamma_s^2&=&-\frac{\sin^2\!\theta}{l^2}\Bigl[(\rho^2+l^2)dt^2-2l\rho^2dtd\varphi\Bigr]\nonumber\\
&&+\frac{\cos^2\!\theta}{\Xi_b}\Bigl[(r^2+b^2)d\psi^2-\frac{\rho^2dt^2}{l^2}\Bigr]\nn\\
&=&-\frac{\sin^2\!\theta}{l^2}\Bigl[(\rho^2+l^2)dt^2-2l\rho^2dtd\varphi\Bigr]\nonumber\\
&&+\frac{\cos^2\!\theta}{\Xi_b}\Bigl[(r^2+b^2)d\psi^2_R+\frac{2b}{l^2}(r^2+b^2)dtd\psi_R\nonumber\\
&&\ \ \ \ -\frac{dt^2}{l^2}\bigl(\rho^2-(r^2+b^2)\frac{b^2}{l^2}\bigr)\Bigr]\,.
\ea
So we get the doubly-spinning charged super-entropic black hole metric
\ba\label{superdouble}
ds^2&=& d\gamma^2_s - \frac{2q \nu_s \omega_s}{\Sigma}+\frac{f \omega_s^2}{\Sigma^2} + \frac{\Sigma dr^2}{\Delta}
+ \frac{\Sigma d\theta^2}{\Xi_b \sin^2\!\theta}\,,\quad
\nn\\
A &=& \frac{\sqrt{3}q\omega_s}{\Sigma}\,,
\ea
where $\nu_s$, $\omega_s$, and $d\gamma^2_s$ are given by \eqref{ho1}--\eqref{ho2},
coordinate $\varphi$ is identified with period $\mu$, $\varphi\sim \varphi+\mu$, and
\ba
\Delta &=& \frac{\rho^4(r^2+b^2)/l^2+q^2+2lbq}{r^2}-2m\,,
\nn\\
f &=& 2m\Sigma -q^2+\frac{2bq}{l}\Sigma\,,\quad
\Xi_b = 1-\frac{b^2}{l^2}\,,
\nn\\
\Sigma &=& r^2+l^2\cos^2\theta + b^2\sin^2\theta\,.
\ea
One can show that this metric satisfies the Einstein--Maxwell-AdS equations. Horizons exist provided $\Delta'(r_+) > 0$.

Note that the super-entropic limit in the $\psi$ (instead of $\phi$) direction would be exactly analogous. However, once the super-entropic limit in the $\phi$-direction is taken, it is no longer possible to perform an additional  $b\to l$ ($\psi$-direction) super-entropic limit. This is because of the $1/\Xi_b$ factor in the $g_{\theta\theta}$ component of the
super-entropic metric \eqref{superdouble}---the corresponding divergence cannot be absorbed into a new azimuthal coordinate.
So we conclude that it is not possible to take successively super-entropic limits in several directions. Neither does it seem possible to set several rotation parameters equal and then perform simultaneously the super-entropic limit in all such directions. What is, however, possible is to combine the super-entropic limit in one direction with the hyperboloid membrane limit in another direction; we discuss this in Appendix ~B.

\subsection{Basic properties}
We now turn to a brief discussion of the horizon and extended thermodynamics of the obtained charged black hole solution. For concreteness we discuss these in coordinates $(t,\varphi,\psi,r,\theta)$, where the coordinate $\psi$ does not rotate at infinity.
\begin{figure*}
\centering
\begin{tabular}{ccc}
 \includegraphics[width=0.32\textwidth,height=0.32\textheight]{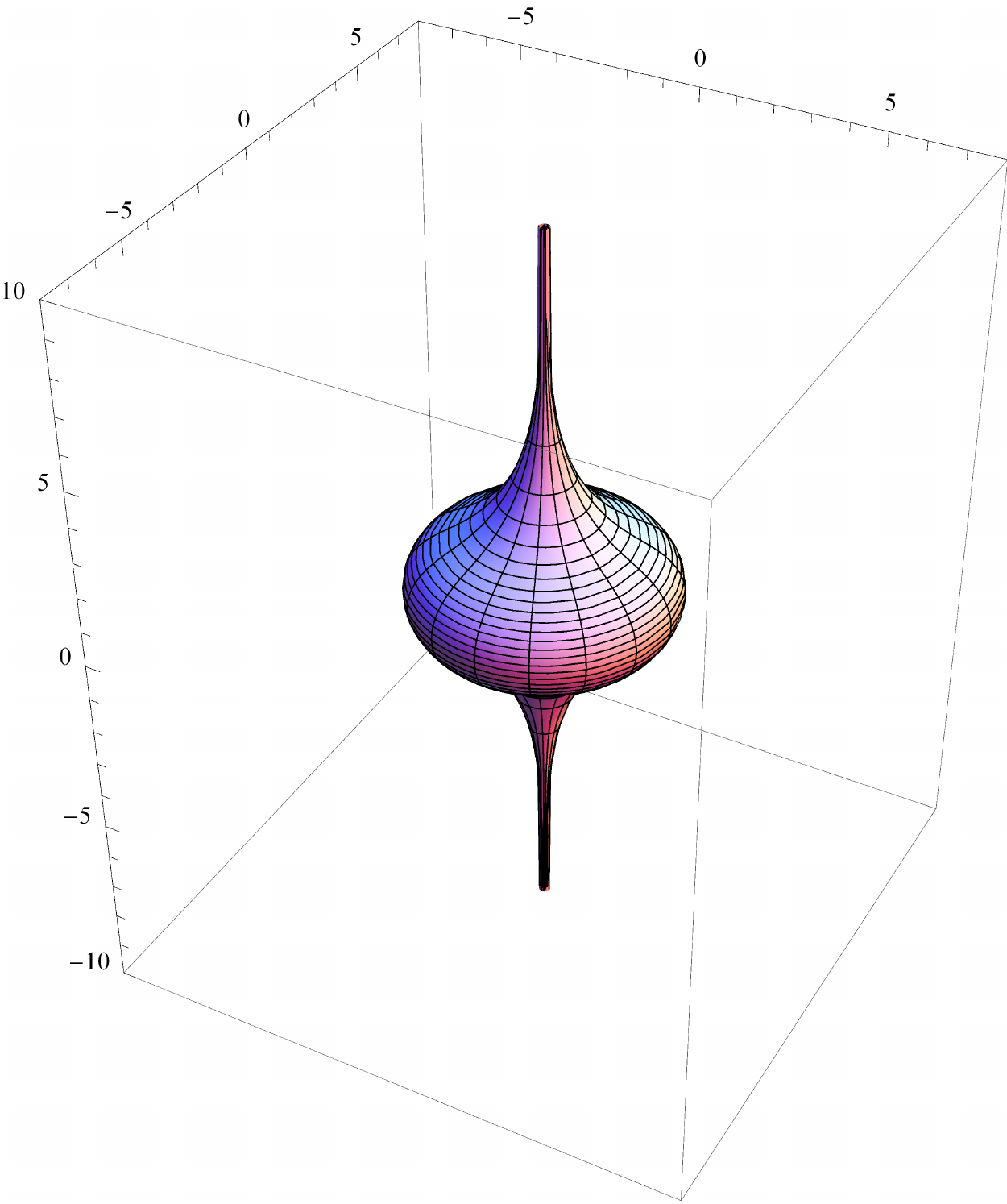} &
 \includegraphics[width=0.32\textwidth,height=0.32\textheight]{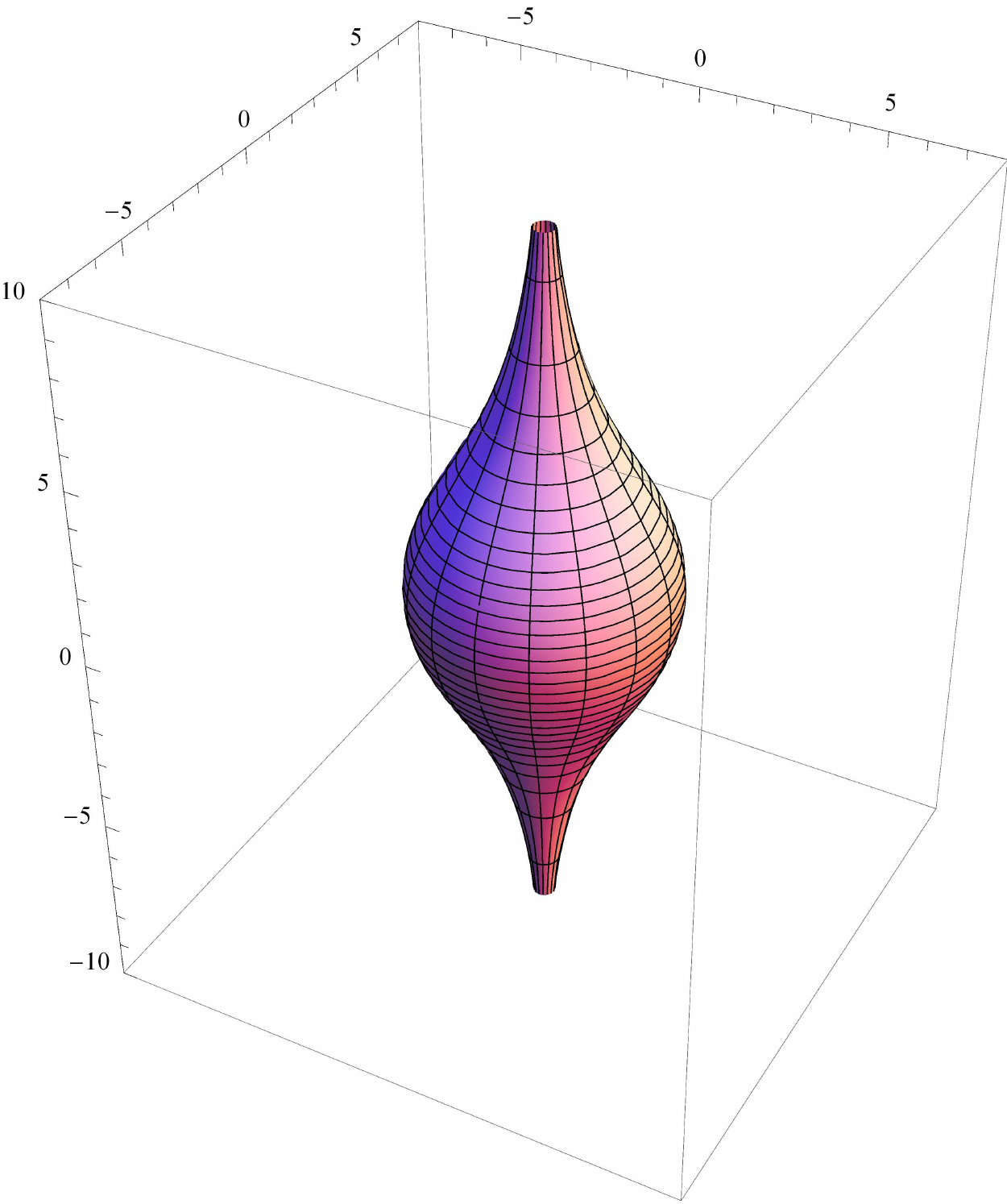}&
 \includegraphics[width=0.32\textwidth,height=0.32\textheight]{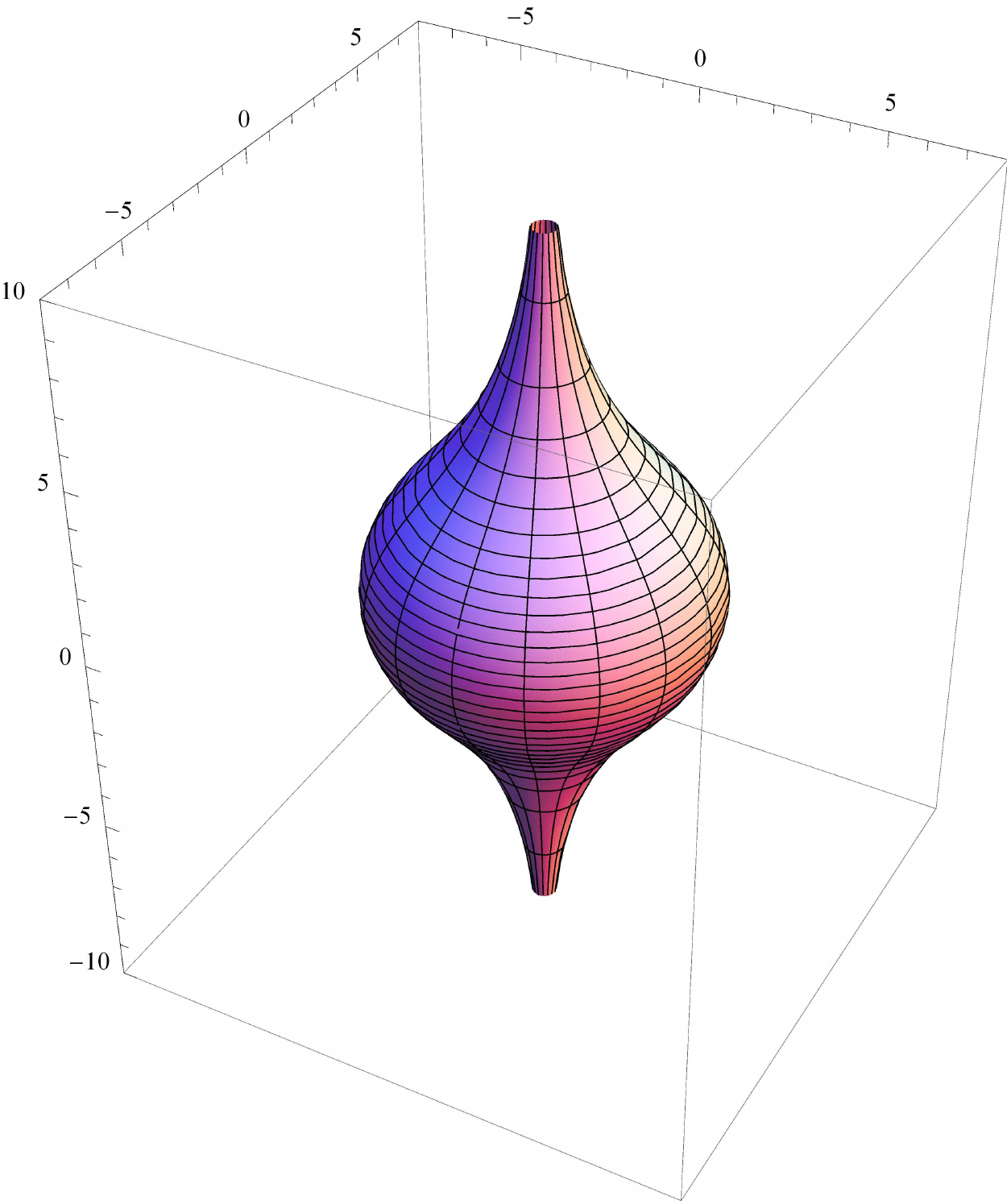}\\
\end{tabular}
\caption{{\bf Horizon embeddings in 5d}. Diagrams display the 2-dimensional $\psi=const.$ horizon slices embedded in $\mathbb{E}^3$ for the following choice of parameters: $b=0,q=0$ (left), $b=0.8\,, \, q=0$ (middle), and $b=0.8,q=45$ (right).  In all plots we have set $\mu = 2\pi, \, r_+ = \sqrt{10}$ and $l=1$. }\label{horizonplots}
\end{figure*}

The induced metric on the horizon takes the form
\ba
ds^2_h&=&\frac{\cos^2\!\theta(r_+^2+b^2)}{\Xi_b}d\psi^2
+\frac{f_+}{\Sigma_+^2}\Bigl(l\sin^2\!\theta d\varphi+\frac{b\cos^2\!\theta d\psi}{\Xi_b}\Bigr)^2
\nonumber\\
&&+\frac{\Sigma_+ d\theta^2}{\Xi_b \sin^2\!\theta}
+\frac{2ql\cos^2\!\theta d\psi}{\Sigma_+}\Bigl(l\sin^2\!\theta d\varphi+\frac{b\cos^2\!\theta d\psi}{\Xi_b}\Bigr)\,,\nonumber\\
\ea
where all quantities are evaluated at $r=r_+$ given by the largest root of $\Delta(r_+)=0$.
For $b=0$ and $q=0$, this reduces to the case studied in the previous section.

Let us for simplicity set $q=0$ and examine the behavior close to the pole $\theta=0$.
As before, we perform the change of coordinates,
\be
\kappa = l(1-\cos\theta)
\ee
and consider the limit $\kappa \to 0$.  The metric becomes,
\ba
ds_h^2 &=& \frac{\rho_+^2}{\Xi_b}\left[\frac{d\kappa^2}{4 \kappa^2 } + \frac{4 \kappa^2 (r_+^2 + b^2)\Xi_b}{l^2 r_+^2}d\varphi^2 \right. \nn\\
&+&  \left. \frac{4b \kappa(r_+^2 + b^2)}{l^2r_+^2}d\varphi d\psi\right] +   \frac{(r_+^2 + b^2)^2}{r_+^2 \Xi_b^2}d\psi^2\,.\qquad
\ea
In particular, the $\psi =$ constant slice reduces to a metric of constant negative curvature on a quotient of the hyperbolic space $\mathbb{H}^2$ showing that the horizon is non-compact.
The slice of constant $(\theta, \varphi)$ is just $S^1$.


The embedding procedure for constant $\psi$ slices of the horizon proceeds as before---the results are
shown in Fig.~\ref{horizonplots}, where we have displayed them for $\mu = 2\pi$, $l=1$, $r_+=\sqrt{10}$ and for various values of $b$ and $q$.  These 2-dimensional slices are visually similar to those of the metric (\ref{sshorizonLimit}): the function $z(\theta) \to \infty$ for $\theta \to \pi, 0$.
Decreasing $b$ results in ``squashing" the horizon, while an increase in the charge parameter causes the horizon to ``bulge out".

The obtained solution is characterized by the following thermodynamic quantities:
\begin{eqnarray}\label{thermoQuantities}
M&=&\frac{\mu}{8} \frac{(m+bq/l)(2+\Xi_b)}{\Xi_b^2} \,, \nonumber\\
J_\varphi&=&\frac{\mu}{4}\frac{lm + bq}{\Xi_b}\,,\quad
J_\psi= \frac{\mu}{8}\frac{2bm + q(b^2+l^2)/l}{\Xi_b^2}\,,\nonumber\\
\Omega_\varphi&=&\frac{l(b^2+r_+^2) + bq}{\rho_+^2(b^2+r_+^2)+lbq}\,,\quad
\Omega_\psi = \frac{b\rho_+^4/l^2 +ql}{\rho_+^2(b^2+r_+^2) +lbq}\,,
\nonumber\\
T &=& \frac{r_+^4\left[2+ (2r_+^2+b^2)/l^2 \right] -(bl+q)^2}{2\pi r_+ \bigl[\rho_+^2(b^2+r_+^2)+lbq \bigr]}\,,
\nn\\
S&=& \frac{\mu \pi \left[(b^2+r_+^2)\rho^2_+ + blq \right]}{4r_+ \Xi_b}=\frac{A}{4}\,,
\nn\\
\Phi &=& \frac{\sqrt{3}qr_+^2}{(b^2+r_+^2)\rho^2_+ + blq}\,,\quad
Q = \frac{\mu \sqrt{3}q}{8 \Xi_b}\,.\qquad
\end{eqnarray}
To calculate the mass and angular momenta, the technique of conformal completion \cite{Ashtekar:1984zz, Ashtekar:1999jx, Das:2000cu} was employed using the Killing vectors $\partial_t, \partial_\varphi$ and $\partial_\psi$. The electric potential is given by $\Phi = \ell^\nu {A}_\nu$ where $\ell^\nu$ is the null generator of the horizon.  The electric charge was computed using Gauss' law, $Q = (1/16\pi) \oint ( \star F - F \wedge A /\sqrt{3})$.

Identifying the mass as the enthalpy of the spacetime, one finds that the extended first law \eqref{eq:first-law} holds if
\ba
V &=& \frac{\mu \pi }{12 r_+^2 l^2 \Xi_b^2}\Big(\bigl[(b^2+3r_+^2)l^2-2b^2r_+^2\bigr]\rho^2_+(b^2+r_+^2)  \nonumber\\
   &+& qbl\left[(2b^2+3r_+^2)l^2 + lbq -b^2r_+^2 \right]  \Big)
\ea
is identified as the thermodynamic volume.  These definitions are also found to satisfy the Smarr relation \eqref{Smarr}. Furthermore, the thermodynamic quantities are found to reduce to those presented earlier for the case of a singly-spinning 5-dimensional super-entropic black hole in the limit $b\to0,\, q\to 0$.
\begin{figure*}
\centering
\begin{tabular}{cc}
\includegraphics[width=0.45\textwidth,height=0.35\textheight]{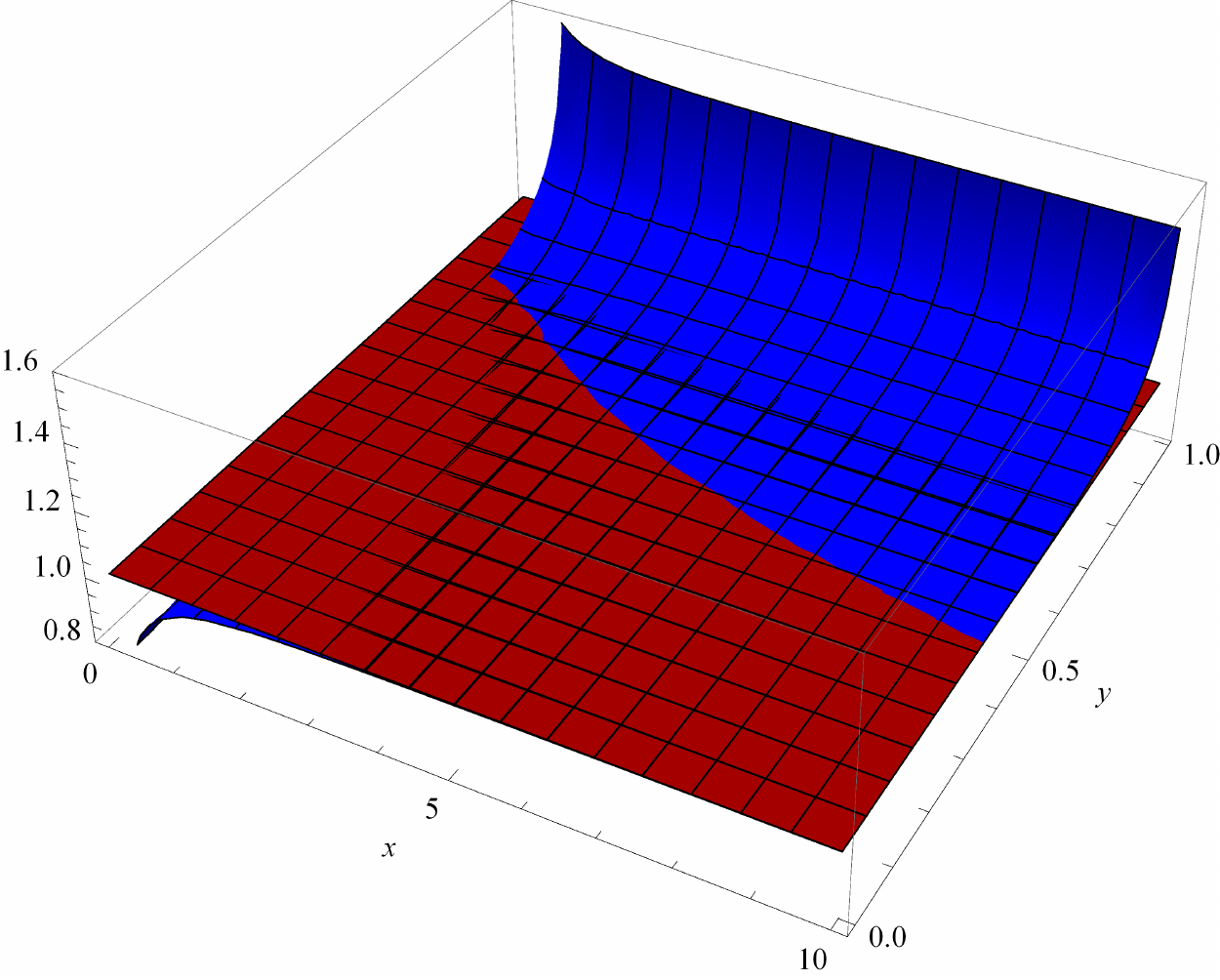}&
\includegraphics[width=0.45\textwidth,height=0.35\textheight]{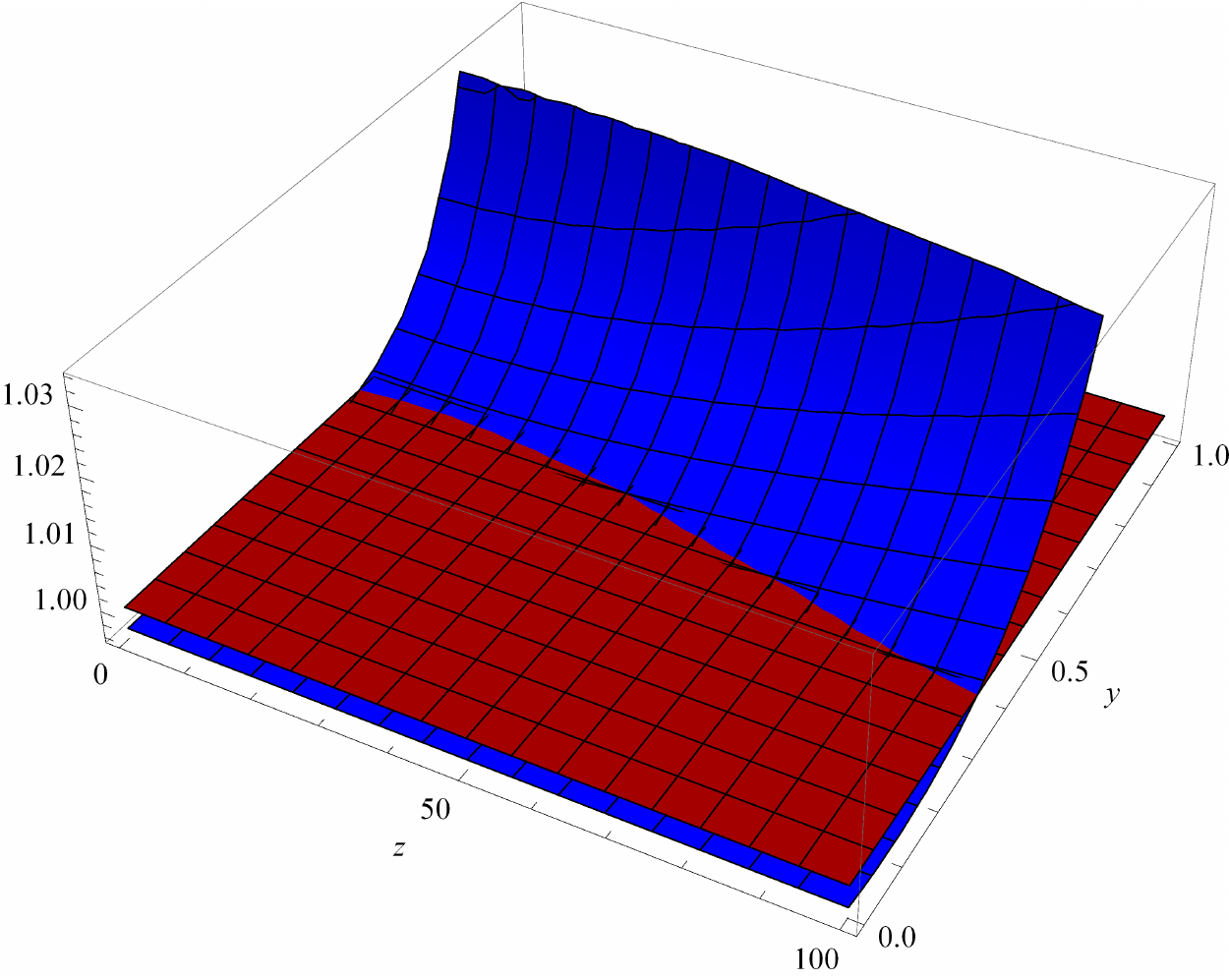}\\
\end{tabular}
\caption{{\bf Isoperimetric ratio.} {\em Left.} A plot of $\mathcal{R}$ (curved blue sheet) for $q=0$; note that a small region of $x$ is excluded due to condition \eqref{5d_horizon_criteria}.
{\em Right.}  A plot of $\mathcal{R}$ for $x=0.5$.
}\label{ipeNoCharge}
\end{figure*}

Let us now discuss the fate of the reverse isoperimetric inequality for this black hole.
In the case where $b=q=0$ we recover the $5$-dimensional singly rotating solution discussed earlier, which  we know to be super-entropic.  Here we find that for non-zero $b,q$ this result is not strictly true: these black holes are super-entropic for some range of $q$ and $b$, but not for all possible $q$ and $b$.

Consider first the case where $b=0$ and $q \not=0$.  In this circumstance, the isoperimetric ratio \eqref{eq:ipe-ratio} is given by,
\be
\mathcal{R}_{b=0} = \left( \frac{r_+^2}{l^2 + r_+^2} \right)^{1/12} < 1\,.
\ee
This indicates that, when $b=0$ these black holes violate the Reverse Isoperimetric Inequality,  satisfying the Isoperimetric Inequality instead.

However, for non-zero $b$ the situation is more complicated.  To study this case we will employ the dimensionless parameters $x = r_+/l$, $y=b/l$ and $z=q/l^2$.  In terms of these parameters, the criteria for the existence of black holes (namely $\Delta'(r_+)>0$) takes the form,
\be\label{5d_horizon_criteria}
x^4(2+2x^2+y^2)\geq (y+z)^2\,
\ee
and we shall enforce this in what follows.
 The simplest case one can consider is the case of nonzero $b$,  but vanishing $q$.  In this case we have,
\be\label{5dratio}
\mathcal{R}^{12}_{q=0} = \left(\frac{1}{27}\right) \frac{(3x^2 + y^2 - 2x^2 y^2)^3}{x^2(1-y^2)^2(x^2+1)(x^2+y^2)}\,.
\ee
This relationship suggests that, for some $y$, the Reverse Isoperimetric Inequality holds.  For example, the factor $1-y^2$ in the denominator ensures that as $y \to 1$, $\mathcal{R} \to \infty$.  However, the Reverse Isoperimetric Inequality does not strictly hold for these black holes.  To see one example of this, consider eq.~(\ref{5dratio}) for small values of $y$. In this case we can write,
\be
\mathcal{R}^{12} = \frac{x^2}{x^2+1}\left(1+ 4y^2 \right) + \mathcal{O}(y^4)\,,
\ee
which, for example, is is less than one when $y^2 < 1/(4x^2)$.  So we see that the Reverse Isoperimetric Inequality is not saved by the addition of another rotation, provided this rotation is sufficiently small. Fig.~\ref{ipeNoCharge} (left) shows a plot of $\mathcal{R}$ (with condition \eqref{5d_horizon_criteria} enforced) for the case $q=0$ highlighting the effects just discussed.

The situation is very similar when charge is included.   The additional effect of the charge can be understood in the following way: for a given value of $x$ the minimum value of $y$ for which $\mathcal{R} > 1$ decreases as $z$ increases.  This is illustrated in Fig.~\ref{ipeNoCharge} (right).

To summarize, not all newly constructed ultraspinning charged AdS black holes violate the
Reverse Isoperimetric Inequality (and so are not `super-entropic'). Depending  on the value of parameters $q$ and $b$, some of them do and some of them do not satisfy $\mathcal{R} \ge 1$. However we shall continue to refer to this entire class of black holes as super-entropic since there
is always a  range of parameters in the metric \eqref{superdouble}
for which the Reverse Isoperimetric Inequality is violated.

\section{General Kerr-AdS black holes}
\label{Sec4}

\subsection{Super-entropic limit}

In this section we shall apply the {super-entropic limit} to the general multi-spinning Kerr-AdS black hole spacetimes \cite{GibbonsEtal:2004, GibbonsEtal:2005}.
Such spacetimes generalize the $d$-dimensional asymptotically-flat rotating black hole spacetimes of Myers and Perry \cite{MyersPerry:1986} and represent the most general
vacuum with cosmological constant,
\be
R_{ab} =-\frac{1}{l^2}(d-1)g_{ab}\,,
\ee
black hole spacetimes of spherical horizon topology without NUT charges \cite{Chen:2006xh}.
In the generalized Boyer--Lindquist coordinates the metric takes the following form:
\be \label{metric}
ds^2=d\gamma^2+\frac{2m}{U} \omega^2+\frac{U dr^2}{F-2m}+d\Omega^2\,,
\ee
where, following the 5-dimensional case,  we have defined
\ba
d\gamma^2&=&-\frac{W\rho^2}{l^2}dt ^2+ \sum_{i=1}^{N} \frac{r^2+a_i^2}{\Xi _i} \mu_i ^2 d\phi _i^2\,,\nn\\
d\Omega^2&=&\sum_{i=1}^{N+\varepsilon}\frac{r^2+a_i ^2}{\Xi _i} d\mu _i ^2
-\frac{1}{W\rho^2}\Bigl(\sum_{i=1}^{N+\varepsilon}\frac{r^2+a_i ^2}{\Xi _i} \mu_i d\mu_i\Bigr)^2\,,\nn\\
\omega&=&W dt -\sum_{i=1}^{N} \frac{a_i \mu_i ^2 d\phi _i}{\Xi _i}\,,
\ea
and, as usual $\rho^2=r^2+l^2$, while
\ba\label{metricfunctions}
W&=&\sum_{i=1}^{N+\varepsilon}\frac{\mu _i^2}{\Xi _i}\,,\quad U=r^\varepsilon \sum_{i=1}^{N+\varepsilon} \frac{\mu _i^2}{r^2+a_i^2} \prod _j ^N (r^2+a_j^2)\,,\nonumber\\
F&=&\frac{r^{\varepsilon -2}\rho^2}{l^2}\prod_{i=1}^N (r^2+a_i^2)\,,\quad \Xi_i=1-\frac{a_i^2}{l^2}\,.\qquad
\ea
To treat even ($\varepsilon=1)$  odd ($\varepsilon=0)$ spacetime dimensionality $d$ simultaneously, we have parametrized
\be
d=2N + 1 + \varepsilon\,
\ee
and in even dimensions set for convenience $a_{N+1}=0$.
The  coordinates $\mu_i$ are not independent, but obey the following constraint:
\begin{equation}\label{constraint}
\sum_{i=1}^{N+\varepsilon}\mu_i^2=1\,.
\end{equation}
In general the spacetime admits $N$ independent angular momenta $J_i$, described by
$N$ rotation parameters $a_i$. Namely, the mass $M$, the angular momenta $J_i$, and the angular velocities of the horizon $\Omega_i$ read \cite{Gibbons:2004ai}
\ba \label{TD}
M &=&\frac{m \omega _{d-2}}{4\pi (\prod_j \Xi_j)}\bigl(\sum_{i=1}^{N}{\frac{1}{\Xi_i}-\frac{1-\varepsilon }{2}}\bigr)\,, \nonumber \\
\quad
J_i &=& \frac{a_i m \omega _{d-2}}{4\pi \Xi_i (\prod_j \Xi_j)}\,,\quad \Omega_i=\frac{a_i (1+\frac{r_+^2}{l^2})}{r_+^2+a_i^2}\,,
\ea
while the temperature $T$, the horizon area $A$, and the entropy $S$ are given by
\ba\label{TS}
T&=&\frac{1}{2\pi }\Bigr[r_+\Bigl(\frac{r_+^2}{l^2}+1\Bigr)
\sum_{i=1}^{N} \frac{1}{a_i^2+r_+^2}-\frac{1}{r_+}
\Bigl(\frac{1}{2}-\frac{r_+^2}{2l^2}\Bigr)^{\!\varepsilon}\,\Bigr]\,,\nonumber\\
A&=&\frac{\omega _{d-2}}{r_+^{1-\varepsilon}}\prod_{i=1}^N
\frac{a_i^2+r_+^2}{\Xi_i}\,,\quad S=\frac{A}{4}\,.
\ea
The horizon radius $r_+$ is determined as the largest root of $F-2m=0$ and $\omega_{d}$ is given by Eq.~\eqref{omega}.

Our goal is to take the super-entropic limit of these spacetimes. Similar to the doubly-spinning 5-dimensional case, it is possible to take
this limit only in one direction, which we choose to be that of the $\phi_j$ 2-plane.

Let us start by looking at the $d\Omega^2$ part of the metric. The trick is to `separate' the $\phi_j$ direction and take the limit $a_j\to l$, the important observation being that
\be
W\Xi_j\to \mu_j^2\,.
\ee
We have
\ba
d\Omega^2&=&
\sum_{i\neq j}\frac{r^2+a_i ^2}{\Xi_i} d\mu _i ^2
-\frac{\Xi_j}{W\Xi_j\rho^2}\Bigl(\sum_{i\neq j}\frac{r^2+a_i ^2}{\Xi _i} \mu_i d\mu_i\Bigr)^2\nn\\
&&-\frac{2(r^2+a_j^2)}{W\Xi_j\rho^2}\mu_jd\mu_j\Bigl(\sum_{i\neq j}\frac{r^2+a_i ^2}{\Xi _i} \mu_i d\mu_i\Bigr)\nn\\
&&+\frac{r^2+a_j ^2}{\Xi_j} d\mu_j ^2\Bigl(1-\frac{r^2+a_j^2}{W\Xi_j\rho^2}\mu_j^2\Bigr)\,.
\ea
Obviously, the limit can be straightforwardly taken for the first 3 terms, while the last term reads
\ba
\frac{r^2+a_j ^2}{W\Xi_j} d\mu_j ^2\!\!\!&&\!\!\!\Bigl(W-\frac{r^2+a_j^2}{\Xi_j\rho^2}\mu_j^2\Bigr)\nn\\
&=&\frac{r^2+a_j ^2}{W\Xi_j} d\mu_j ^2\Bigl(\sum_i\frac{\mu_i^2}{\Xi_i}-\frac{\rho^2-\Xi_j l^2}{\Xi_j\rho^2}\mu_j^2\Bigr)\nn\qquad\\
&=&\frac{r^2+a_j ^2}{W\Xi_j} d\mu_j ^2\Bigl(\sum_{i\neq j}\frac{\mu_i^2}{\Xi_i}+\frac{l^2}{\rho^2}\mu_j^2\Bigr)\,.
\ea
Putting everything together, we have $d\Omega^2\to d\Omega^2_s$, where
\ba\label{Omegas}
d\Omega^2_s&=&
\sum_{i\neq j}\frac{r^2+  a^2_i }{\Xi _i} d\mu _i ^2 -2\frac{ d\mu_j}{\mu_j}\Bigl(\sum_{i\neq j}^{N+\varepsilon}\frac{r^2+  a^2_i }{\Xi _i} \mu_i d\mu_i\Bigr) \nonumber\\
&&+ \frac{d\mu _j ^2}{\mu _j ^2}\Bigl( \rho^2\hat W+l^2 \mu _j ^2 \Bigr)\,,
\ea
where we have defined
\be\label{hatW}
\hat W=\sum_{i\neq j}\frac{\mu_i^2}{\Xi_i}\,.
\ee
Note that there is no further scope for setting any other $a_i\to l$ --- this additional limit will cause the preceding expression to diverge.

To proceed further, we switch to the rotating frame by setting
\be\label{azirescale2}
\phi_j=\phi_{j}^R+\frac{a_j}{l^2}t\,.
\ee
Then we have
\be
\omega=(\hat{W}+\mu^2_j) dt -\frac{a_j\mu_j^2d\phi_{j}^R}{\Xi_j}-\sum_{i\neq j} \frac{a_i \mu_i ^2 d\phi_i}{\Xi _i}\,.
\ee
Upon setting
\be
\varphi_j=\frac{\phi_{j}^R}{\Xi_j}\,,
\ee
and taking the limit $a_j\to l$, we have $\omega\to \omega_s$,
\be\label{omegas}
\omega_s=(\hat{W}+\mu^2_j) dt -l\mu_j^2d\varphi_{j}-\sum_{i\neq j} \frac{a_i \mu_i ^2 d\phi_i}{\Xi _i}\,.
\ee
Similarly we have
\ba
d\gamma^2&=&\bigl(-W\rho^2+\frac{a_j^2}{l^2}\frac{\mu_j^2}{\Xi_j}(r^2+a_j^2)\bigr)\frac{dt^2}{l^2}+\frac{r^2\!+\!a_j^2}{\Xi_j}\mu_j^2(d\phi_{j}^R)^2\nn\\
&+&2a_j\mu_j^2(r^2+a_j^2)\frac{dt d\phi_{j}^R}{l^2\Xi_j}
+\sum_{i\neq j}\frac{r^2+a_i^2}{\Xi_i}\mu_i^2d\phi_{i}^2\,,\qquad
\ea
which limits to
\ba\label{gammas}
d\gamma^2_s&=&-\Bigl((\hat W+\mu_j^2)\rho^2+\mu_j^2l^2\Bigr)\frac{dt^2}{l^2}+\frac{2 \rho^2\mu_j^2 dt d\varphi}{l}\nn\\
&&+\sum_{i\neq j}\frac{r^2+a_i^2}{\Xi_i}\mu_i^2d\phi_{i}^2\,.
\ea
Putting everything together we arrive at multiply spinning super-entropic black holes, given by
\be\label{multiSuperentropic}
ds^2=d\gamma^2_s+\frac{2m}{U} \omega_s^2+\frac{U dr^2}{F-2m}+d\Omega_s^2\,,
\ee
where $d\Omega_s^2$, $\hat W$, $\omega_s$, $d\gamma_s^2$ are given by \eqref{Omegas}, \eqref{hatW}, \eqref{omegas}, \eqref{gammas} and we have
\ba\label{metricfunctionsS}
U&=&r^\varepsilon \Bigl(\mu_j^2+\sum_{i\neq j}\frac{\mu _i^2\rho^2}{r^2+a_i^2}\Bigr)\prod_{k\neq j}^N (r^2+a_k^2)\,,\\
F&=&\frac{r^{\varepsilon -2}\rho^4}{l^2}\prod_{i\neq j}^N (r^2+a_i^2)\,,\quad \Xi_i=1-\frac{a_i^2}{l^2}\  \mbox{for}\  i\neq j\,.\qquad\nn
\ea
We stress that the $\mu$'s are not independent as they still satisfy the constraint \eqref{constraint}. We also identify the coordinate
$\varphi_j\sim \varphi_j+\mu$.

\subsection{Basic properties}
Let us briefly discuss some of the basic properties of the newly constructed multispinning super-entropic black holes.

First, the super-entropic geometry inherits from the Kerr-AdS spacetimes a remarkable property---it possesses a hidden symmetry associated with the principal Killing--Yano tensor, $h=db$ \cite{KubiznakFrolov:2007}. After the transformation \eqref{azirescale2}, the Killing--Yano potential in the original Kerr-AdS spacetime reads (cf. Eq. (B.14) in \cite{Kubiznak:2008qp})
\ba
2b&=&\Bigl(r^2+a_j^2\mu_j^2+\sum_{i\neq j}a_i^2\mu_i^2\bigl[1+\frac{r^2+a_i^2}{l^2\Xi_i}\bigr]\Bigr)dt\nn\\
&&-a_j\mu_j^2\frac{r^2+a_j^2}{\Xi_j}d\phi_j^R-\sum_{i\neq j}a_i\mu_i^2\frac{r^2+a_i^2}{\Xi_i}d\phi_i\,.\qquad
\ea
After the ultraspinning limit, this gives a Killing--Yano potential for the super-entropic black holes \eqref{multiSuperentropic}, given by
\ba
2b_s&=&\Bigl(r^2+l^2\mu_j^2+\sum_{i\neq j}a_i^2\mu_i^2\bigl[1+\frac{r^2+a_i^2}{l^2\Xi_i}\bigr]\Bigr)dt\nn\\
&&-l\mu_j^2\rho^2d\varphi_j-\sum_{i\neq j}a_i\mu_i^2\frac{r^2+a_i^2}{\Xi_i}d\phi_i\,.\qquad
\ea
Similar to the original Kerr-AdS spacetimes, such a tensor guarantees complete integrability of geodesic motion as well as separability of
 various test field equations in these spacetimes \cite{Frolov:2008jr}. In particular, this implies that
 one can study geodesic completeness in a way similar to what we did in Sec.~\ref{geodesics} for 4-dimensional super-entropic black holes.

Let us now turn to the horizon. The corresponding induced metric reads
\ba\label{multisupHorizon}
ds_h^2&=&
\frac{F}{U} \Bigl( l \mu_j^2 d\varphi_j + \sum_{i\neq j}^{N} \frac{a_i \mu_i ^2 d\phi_i}{\Xi _i}\Bigr)^2\nn\\
&&+ \sum_{i\neq j}^{N} \frac{r_+^2+a_i^2}{\Xi _i} \mu_i ^2 d\phi _i^2+d\Omega_s^2\,,
\ea
where all the quantities are evaluated at $r=r_+$. As before, we are interested in the behavior near $\mu_j = 0$.
Considering the $\phi_i=const.$ and $\mu_i=const.$ slice, and in the limit $\mu_j\to 0$, the previous expression has the following leading order expansion:
\be
ds_h^2\approx \rho^2\hat W\Bigl(\frac{d\mu_j^2}{\mu_j^2}+\frac{Fl^2}{U\rho^2\hat W}\mu_j^{4}d\varphi_j^2\Bigr)\,.
\ee
As before, this is a metric of constant negative curvature on $\mathbb{H}^2$ and so the super-entropic limit has yielded a non-compact horizon  here as well.

Finally, we shall discuss the thermodynamics of the obtained solution. Employing the same technique as before, we recover the following thermodynamic quantities:
\ba
M &=& \frac{m \omega_{d-2}}{4 \pi  \prod_{k \not= j} \Xi_k} \Bigl(\sum_{i\not= j} \frac{1}{\Xi_i} + \frac{1+\varepsilon}{2} \Bigr)\,,
\nn\\
\Omega_{j} &=& \frac{l}{\rho_+^2}\,, \quad \Omega_{i\not=j} = \frac{a(l^2+r_+^2)}{l^2(r_+^2 + a_i^2)}\,,
\nn\\
J_{j} &=& \frac{l m \omega_{d-2}}{4 \pi \prod_{k\not= j} \Xi_k}\,,
\quad J_{i \not= j} = \frac{a_i m \omega_{d-2}}{4 \pi \Xi_i \prod_{k\not=j} \Xi_k }\,,
\nn\\
T&=& \frac{1}{2 \pi} \Bigl[\frac{r_+}{l^2}\Bigl(1 + \sum_{i\not= j}^N \frac{\rho_+^2}{a_i^2\!+\!r_+^2} \Bigr)
-\frac{1}{r_+} \Bigl(\frac{1}{2} - \frac{r_+^2}{2l^2} \Bigr)^\varepsilon \Bigr]\,,
\nn\\
A &=& \frac{\omega_{d-2}}{r_+^{1-\varepsilon}}\rho_+^2 \prod_{i\not= j}^N \frac{r_+^2 + a_i^2}{\Xi_i}\,, \quad S= \frac{A}{4}\,.
\ea
One can verify that these satisfy the traditional first law.  If we now identify the black hole mass as the enthalpy of the spacetime, we find that the extended first law is satisfied provided the thermodynamic volume is given by
\be\label{manyrotsvol}
V = \frac{r_+ A}{d-1} + \frac{8 \pi }{(d-1)(d-2)} \sum_{i\not=j} a_i J_i\,.
\ee
It is easy to see that, in the case of a single rotation, Eq.~\eqref{manyrotsvol} reduces to the expression presented in \eqref{singlespin} for the thermodynamic
volume---namely the naive geometric volume.  This allows us to make some immediate conclusions regarding the Reverse Isoperimetric Inequality.   We see that for small rotation parameters $a_i$, the thermodynamic volume will be close to the naive geometric volume, $V_0 = r_+A/(d-1)$.  As was discussed in the context of the singly spinning super-entropic black hole, the naive geometric volume satisfies the Isoperimetric Inequality.  Therefore, we can
immediately conclude that in the general case the Reverse Isoperimetric Inequality is violated for small values of the rotation parameters $a_i$.  Thus it is only for some parameter values that these ultraspinning black holes violate $\mathcal{R} \ge 1$ and so are super-entropic, though we shall refer to this entire class by that name.

\section{Conclusions}

We have utilized the novel super-entropic ultraspinning limit to generate a broad new class of black hole solutions,  significantly deepening the analysis performed in
\cite{Gnecchi:2013mja,Klemm:2014rda,Hennigar:2014cfa}. Namely, we have constructed new higher-dimensional multiply spinning super-entropic black holes
starting from the general Kerr-AdS metrics in all dimensions and the general rotating black hole of minimal gauged supergravity in five dimensions. All such new solutions are super-entropic in one direction. It seems impossible to perform the super-entropic limit in several directions, neither successively
for several rotation parameters nor
simultaneously for equal spinning black holes. However we found that it is possible to combine the super-entropic limit in one direction with a hyperboloid membrane limit in another direction, obtaining a novel super-entropic hyperboloid membrane solution of Einstein's equations. The technical aspects of these various limits are discussed in the Appendices.

The super-entropic limit can be thought of as a simple generating procedure. Starting from a known rotating asymptotically AdS black hole solution, one performs
a coordinate transformation that puts the metric into ``rotating at infinity'' coordinates in one azimuthal direction.  This rotation is then boosted to the speed of light by taking the (naively) singular $a \to l$ limit in a sensible way.  The result is a nontrivial change in the structure and topology of the spacetime, since it is no longer possible to return to non-rotating coordinates and the axis of rotation is excised from the spacetime.  In all cases examined, the resulting black holes possess the unique feature of having a non-compact event horizon of finite area.  Topologically, the event horizons are spheres with two punctures (i.e. cylinders), and as such these black holes could be considered the AdS generalization of the asymptotically flat black cylinders considered in \cite{Emparan:2009vd,Armas:2010hz}, though they do not have a smooth flat-space limit.


In the context of extended phase space thermodynamics  the entropy per given thermodynamic volume of all solutions was found to exceed the naive limit set by the conjectured Reverse Isoperimetric Inequality \cite{Cvetic:2010jb}, for at least some range of the parameters.  For this reason we refer to all such black holes as ``super-entropic''.
   This feature is attributed to be a result of the non-compact horizons of these black holes.  As suggested in \cite{Hennigar:2014cfa}, these super-entropic black holes do not necessarily spell the end of the Reverse Isoperimetric Inequality Conjecture, but rather suggest that it applies to black holes with compact horizon, in analogy to the standard geometric isoperimetric inequality in Euclidean space.  The proof of this (restricted) conjecture remains an interesting open problem.

\section*{Acknowledgments}

We thank R. Gregory for helpful discussions.
This research was supported in part by Perimeter Institute for Theoretical Physics and by the Natural Sciences and Engineering Research Council of Canada. Research at Perimeter Institute is supported by the Government of Canada through Industry Canada and by the Province of Ontario through the Ministry of Research and Innovation.

\appendix

\section{Black Brane Limit} \label{AppA}
Ultraspinning black holes were first studied by Emparan and Myers  \cite{Emparan:2003sy} in an analysis focusing on the stability of Myers--Perry black holes \cite{Myers:1986un} in the limit of large angular momentum. As briefly discussed in the introduction, for AdS black holes several physically distinct ultraspinning limits are possible. In this appendix we review the first type---the black brane ultraspinning limit---first studied by Caldarelli et al. \cite{Caldarelli:2008pz}
for Kerr-AdS black holes. The procedure consists of taking a limit where one or more rotation parameters, $a_i$, approach the AdS radius, $l$, $a_i\to l$,
keeping the physical mass $M$ of the black hole spacetime fixed while simultaneously zooming in to the pole.
This limit is sensible only for $d \ge 6$ and yields a vacuum solution of Einstein equations (with zero cosmological constant) describing a static black brane.
Armas and Obers later showed that the same solution can be obtained by taking $a \to \infty$ while keeping the ratio $a/l$ fixed, their approach having the advantage of being directly applicable to dS solutions as well \cite{Armas:2010hz}.

In this appendix we follow the original reference \cite{Caldarelli:2008pz} and demonstrate the procedure for the multiply spinning Kerr-AdS black hole
spacetimes discussed in section~\ref{Sec4}. We also comment on an (im)possibility of taking the black brane limit starting from the newly constructed super-entropic black holes.

\subsection{Limit in one direction}
Let us first discuss how to take the black brane limit in one direction, associated with the $j$ 2-plane.
Starting from the Kerr-AdS metric~\eqref{metric} we perform the following scaling:
\be\label{stdult}
t = \epsilon^2 \hat{t}\,, \quad r = \epsilon^2 \hat{r}\,,  \quad  \mu_j =\epsilon^{\frac{d-1}{2}} {\sigma}/{l}\,,
\ee
where
\be
\epsilon=\Xi_j^{\frac{1}{d-5}}\to 0 \quad \mbox{as}\quad a_j\to l\,.
\ee
Since we want to  keep the physical mass $M$ and angular momenta $J_{i}$ finite for all $i$, we have to have $m \sim  \epsilon^{2(d-5)}$. Namely, we set
\be
m \to  \epsilon^{2(d-5)} \hat{m} l^2\,,
\ee
where the factor $l^2$ was chosen to cancel a factor of $l^2$ in $U$ after the rescaling. For the limit to work, we must have also keep $m/U$ finite.  Using the scalings \eqref{stdult},
\ba\label{Uult1}
U&=&r^\varepsilon \sum_{i=1}^{N+\varepsilon} \frac{\mu _i^2}{r^2+a_i^2} \prod _{k} ^N (r^2+a_{k}^2) \\
&=& \epsilon^{2 \varepsilon} \hat{r}^\varepsilon
\Bigl(\frac{\sigma^2}{l^2}\epsilon^{d-1}\!\!
+\!\! \sum_{i\neq j}^{N+\varepsilon} \frac{\mu _i^2 (\epsilon^4 \hat{r}^2+a_j^2)}{\epsilon^4 \hat{r}^2+a_i^2} \Bigr)
\prod _{k\neq j}^N (\epsilon^4 \hat{r}^2+a_k^2)\nonumber \, .
\ea
We see from this that we will not have $U \sim \epsilon^{2(d-5)}\hat{U}$ unless we rescale the $a_i$'s so that
\be
a_i\to \epsilon^2 \hat{a}_i\quad \mbox{for}\quad i\neq j\,.
\ee
Let us define the following two functions for future reference:
\begin{eqnarray}\label{NewFunctions}
	\hat{U}
	&=&
	\hat{r}^\varepsilon
	\Bigl(
		\sum_{i\neq j}^{N+\varepsilon}
		\frac{\mu _i^2}{\hat{r}^2+\hat{a}_i^2}
	\Bigr)
	\prod _{k\neq j}^N ( \hat{r}^2+\hat{a}_k^2)\,,\nn\\
	\hat{F}
	&=&
	\hat{r}^{\varepsilon -2} \prod_{i\neq j}^N (\hat{r}^2+\hat{a}_i^2)\,.
\end{eqnarray}
Then we find
\ba\label{Uult2}
U &=& \epsilon^{4N\!-8+2 \varepsilon} \hat{r}^\varepsilon
\Bigl( \frac{\sigma^2}{l^2} \epsilon^{d+3}\! +\!\!
\sum_{i\neq j}^{N+\varepsilon} \frac{\mu _i^2(\epsilon^4 \hat{r}^2\!+\!a_j^2)
}{\hat{r}^2+\hat{a}_i^2} \Bigr)\prod _{k\neq j}^N ( \hat{r}^2\!+\!\hat{a}_k^2) \nonumber \\
&=& \epsilon^{2(d-5)} a_j^2\hat U+{\cal{O}}(\epsilon^{2d-6})\,,
\ea
giving (in the limit $\epsilon\to 0$)
\begin{eqnarray}
	\frac{m}{U}
	& \sim &
	\frac{
		\epsilon^{2(d-5)} \hat{m} l^2
	}{
		\epsilon^{2(d-5)} a_j^2 \hat{U}
	}
	\,\to\,
	\frac{ \hat{m} }{ \hat{U} }\,.
\end{eqnarray}
Also the limits of
$W$ and $F$ are now easy to take
\ba\label{metricfunctions2}
F&=& r^ {\varepsilon -2} \Bigl(1\!+\!\frac{r^2}{l^2}\Bigr) \prod_{i=1}^N (r^2\!+\!a_i^2)=\epsilon^{2(d-5)} a_j^2 \hat{F} + \mathcal{O}(\epsilon^{2d-6})\,,\nn\\
W&=&\sum_{i=1}^{N+\varepsilon}\frac{\mu _i^2}{\Xi _i} = \epsilon^{4} \frac{\sigma^2}{l^2}
+\sum_{i\neq j}^{N+\varepsilon}\frac{\mu _i^2}{\left(1\!-\!\epsilon^4 \hat{a}^2_i/l^2\right) } \nn\\
&=&\sum_{i\neq j}^{N+\varepsilon}{\mu _i^2}+\mathcal{O}(\epsilon^{4})=1+\mathcal{O}(\epsilon^{4})\,.
\ea
Hence we get the correct scaling of $F$ to keep $U/(F-2m)$ finite.
We also have
\ba
&&\sum_{i=1}^{N} \frac{a_i \mu_i ^2 d\phi _i}{\Xi _i}
= \epsilon^4\frac{a_j \sigma^2 d\phi _j}{l^2}
+ \sum_{i\neq j}^{N} \frac{\epsilon^2 \hat{a}_i \mu_i ^2 d\phi _i}{\left(1-\epsilon^4 \hat{a}^2_i/l^2\right) }  \nonumber\\
&&\quad = \epsilon^2\sum_{i\neq j}^{N} {\hat{a}_i \mu_i ^2 d\phi _i}
+ \mathcal{O}(\epsilon^4)\,, \nonumber\\
&&\sum_{i=1}^{N} \frac{r^2+a_i^2}{\Xi _i} \mu_i ^2 d\phi _i^2 \nonumber\\
&&\quad= \frac{\epsilon^8 r^2+\epsilon^4 a_j^2}{l^2} \sigma^2 d\phi _j^2
+\epsilon^4 \sum_{i\neq j}^{N} \frac{\hat{r}^2+\hat{a}_i^2}{\left(1-\epsilon^4 \hat{a}^2_i/l^2\right)} \mu_i ^2 d\phi _i^2 \qquad\nonumber\\
&&\quad=\epsilon^4 \frac{a_j^2}{l^2} \sigma^2 d\phi _j^2
+\epsilon^4 \sum_{i\neq j}^{N} (\hat{r}^2+\hat{a}_i^2) \mu_i ^2 d\phi _i^2
+ \mathcal{O}(\epsilon^8)\,,
\ea
and the $d\mu_i$ terms give
\ba
&&\sum_{i=1}^{N+\varepsilon}\!\frac{r^2\!+\!a_i ^2}{\Xi _i} d\mu _i ^2
=  \frac{\epsilon^8 \hat{r}^2\!+\!\epsilon^4  a_j ^2}{l^2} d\sigma^2 \!+\!\epsilon^4 \sum_{i\neq j}^{N+\varepsilon}\!\frac{\hat{r}^2+\hat{a}_i ^2}{\left(1\!-\!\epsilon^4 \hat{a}^2_i/l^2\right) } d\mu _i ^2
\nonumber\\
&&\quad=  \epsilon^4 {\frac{a_j^2}{l^2}} d\sigma^2
+\epsilon^4 \sum_{i\neq j}^{N+\varepsilon}(\hat{r}^2+\hat{a}_i ^2)d\mu _i ^2  + \mathcal{O}(\epsilon^8)\,, \nonumber\\
&&\Bigl(\sum_{i=1}^{N+\varepsilon}\frac{r^2+a_i ^2}{\Xi _i} \mu_i d\mu_i\Bigr)^2\nonumber\\
&&\quad =\Bigl(\frac{\epsilon^8 \hat{r}^2+\epsilon^4  a_j ^2}{l^2} \sigma d\sigma
+\epsilon^4 \sum_{i \ne j}^{N+\varepsilon}\frac{\hat{r}^2+\hat{a}_i ^2}{\left(1-\epsilon^4 \hat{a}^2_i/l^2\right) } \mu_i d\mu_i\Bigr)^2\nonumber\\
&&\quad = \mathcal{O}(\epsilon^8)\,.
\ea
Now that we know how all the components of the metric scale at lowest order as we take the black brane ultraspinning limit, we can set $\phi_j = \varphi$, and rescale the metric by a constant conformal factor, $s=\epsilon^2 \hat{s}$. There are no components of order less than 4 in the rescaled metric, so we may cancel the $\epsilon^4$ and complete the limit $a_j \to l$.

The obtained metric is a vacuum solution of Einstein equations with zero cosmological constant that describes a (static in the original 2-plane) black brane
\ba \label{uspinmetric}
d\hat{s}^2
&=&
- d\hat{t}^2
+\frac{2\hat{m}}{ \hat{U}} \Bigl( d\hat{t}- \sum_{i\neq j}^{N} {\hat{a}_i \mu_i ^2 d\phi _i}\Bigr)^2
+ \frac{\hat{U} d\hat{r}^2}{\hat{F}-2\hat{m}}\nn \\
&+&d\sigma^2
+ \sigma^2 d\varphi^2
+ \sum_{i\neq j}^{N+\varepsilon}(\hat{r}^2+\hat{a}_i ^2)d\mu _i ^2
\nonumber \\
&+& \sum_{i\neq j}^{N} (\hat{r}^2+\hat{a}_i^2) \mu_i ^2 d\phi _i^2\,.
\ea
Here, the metric functions $\hat F$ and $\hat U$ are given by  \eqref{NewFunctions}, and the coordinates $\mu_i$
are bound to satisfy the following constraint:
\begin{equation}\label{constraint2}
\sum_{i\neq j}^{N+\varepsilon}\mu_i^2=1\,.
\end{equation}
Note that in the process of taking the black brane limit we have `lost' the AdS radius $l$ and no longer have an asymptotically AdS space.
This is in contrast to the super-entropic and hyperboloid membrane limits which retain their asymptotic AdS structure.
Another difference is that the black brane limit can be simultaneously
taken in several directions\footnote{Since the result of the black brane limit
is no longer AdS it is not possible to take several such limits successively.},
whereas this is impossible for the super-entropic limit. We shall discuss this next.

\subsection{Limit in multiple directions}

The black brane limit, contrary to the super-entropic limit, can be simultaneously taken in several directions \cite{Emparan:2009vd,Armas:2015nea}.
Let us start from the Kerr-AdS solution \eqref{metric} where we set several rotation parameters equal, $a_j=a$ for $j=1,\dots , n$.
We want to take the limit $a\to l$.
The procedure is very similar to the above, but we must choose the various scalings more carefully.
We begin as before: to keep the mass finite, we must now have $m\sim \Xi^{n+1}_j$ from \eqref{TD}.
In this case all the angular momenta $J_j$ also remain finite.
If $r$ and the remaining $a_k$ scale as before, then, after writing $a_j=a$ for $j=1\ldots n$, reindexing $\hat{a}_k$ to $k = 1,\ldots N-n+\varepsilon$,
and defining
\ba\label{UFhat2}
\hat U_n&=& \hat{r}^\varepsilon
\Bigl(\,\sum_{i\neq j}^{N-n+\varepsilon}\!\!  \frac{\mu _i^2}{\hat{r}^2\!+\!\hat{a}_i^2} \Bigr)
\prod _{k=1}^{N-n}  ( \hat{r}^2\!+\!\hat{a}_k^2)\,,\nn\\
\hat{F}_n&=& r^ {\varepsilon -2}\prod_{i =1}^{N-n} (\hat{r}^2+\hat{a}_i^2)\,,
\ea
we have
\ba\label{Uult3}
U &=& \epsilon^{4N-4n+2 \varepsilon} \hat{r}^\varepsilon \prod _{k=1}^{N-n} ( \hat{r}^2+\hat{a}_k^2)
\prod _{k=1}^{n}(\epsilon^4 \hat{r}^2+a_{k}^2)
 \nonumber \\
&&\ \times \Bigl( \sum_{j=1}^{n} \frac{\mu _j^2}{\epsilon^4\hat{r}^2+\hat{a}_j^2} +\frac{1}{\epsilon^4}
\sum_{i\neq j}^{N-n+\varepsilon}
\frac{\mu _i^2}{\hat{r}^2+\hat{a}_i^2} \Bigr) \nonumber\\
&=& \epsilon^{2(d-2n-3)} l^{2n}\hat U_n  +\cdots\,.
\ea
Since we want $m/U_n \sim \epsilon^{0}$, in the multispinning case we must have the following scaling:
\be\label{stdult2}
t = \epsilon^2 \hat{t}\,,
\quad
m=l^{2n}\epsilon^{2(d-2n-3)} \hat{m}\,,
\quad
\epsilon=\Xi_j^{\frac{n+1}{2(d-2n-3)}}\,.
\ee
Choosing further
\begin{eqnarray}
	\mu_j = \epsilon^{\frac{d-1}{n+1}} \sigma_j/l\quad \mbox{for}\quad j=1,\dots, n\,,
\end{eqnarray}
we also have
\ba
W &\to& \sum_{k=n+1}^{N+\varepsilon} \mu_k^2 + \mathcal{O}(\epsilon^4) = 1\,,\nn\\
F&=& r^ {\varepsilon -2} \Bigl(1+\frac{r^2}{l^2}\Bigr) \prod_{i=1}^{N} (r^2+a_i^2) \nonumber\\
&=& \epsilon^{2(d-2n-3)}l^{2n}\hat{F}_n\,.
\ea
The other limits are similar, and so the black brane limit taken in $n$ directions gives the following metric
($s=\epsilon^2 \hat{s}$):
\ba \label{muspinmetric}
d\hat{s}^2
&=&
- d\hat{t}^2
+\frac{2\hat{m}}{ \hat{U}_n} \Bigl( d\hat{t}- \sum_{i\neq j}^{N-n} {\hat{a}_i \mu_i ^2 d\phi _i}\Bigr)^2
+ \frac{\hat{U}_n d\hat{r}^2}{\hat{F}_n-2\hat{m}} \nonumber \\
&+&\sum_{j=1}^n \left(d\sigma_j^2 + \sigma_j^2 d\varphi_j^2\right)
+ \sum_{i=1}^{N-n+\varepsilon}(\hat{r}^2+\hat{a}_i ^2)d\mu _i ^2\nn \\
&+&
\sum_{i=1}^{N-n} (\hat{r}^2+\hat{a}_i^2) \mu_i ^2 d\phi _i^2\,,
\ea
where we have set $\varphi_{i} = \phi_i$ for $i = n+1, \ldots N+\varepsilon$, functions $\hat U$ and $\hat F$ are defined in \eqref{UFhat2}, and the constraint now reads
\begin{equation}\label{constraint2}
	\sum_{i=1}^{N-n+\varepsilon}\mu_i^2=1\, .
\end{equation}
As when this limit is taken in only one direction, it can only be done for $d\ge 6$ and the resulting space is no longer AdS.

Let us finally mention that we were unable to obtain the black brane limit of the super-entropic black holes
\eqref{multiSuperentropic}---the super-entropic and the black brane limits seem incompatible. However, as we discuss in the next appendix,
it is possible to combine the super-entropic limit with the hyperboloid membrane limit.

\section{Hyperboloid membrane limit}

In this appendix, we examine another type of the ultraspinning limit---the hyperboloid membrane limit---and its compatibility with the super-entropic limit.
The hyperboloid membrane limit was first studied in \cite{Caldarelli:2008pz,Caldarelli:2012cm}, where it was found applicable to the Kerr-AdS spacetime for $d \ge 4$.
In this limit, one lets the rotation parameter $a$ approach the AdS radius $l$, $a \to l$, while scaling the polar angle $\theta \to 0$ in a way so that the coordinate $\sigma$ defined by
\begin{align}
	\sin\theta
	&=
	\sqrt{\Xi} \sinh(\sigma/2)
	\label{eqn:hyperboliclimit}
\end{align}
remains fixed.
Contrary to the super-entropic limit, this limit does not require any special rotating frame.
We shall now demonstrate how this works for black holes in four and five dimensions.

In four dimensions, applying the coordinate transformation \eqref{eqn:hyperboliclimit} to the Kerr--Newman-AdS metric \eqref{KNADS2} and taking the limit $a \to l$, we find
\ba
d s^2&=&
	-f\big( dt - l \sinh^2(\sigma/2) \; d \phi \big)^2
	+ \frac{dr^2}{f}\nn\\
&&+ \frac{\rho^2}{4}\big( d \sigma^2 + \sinh^2\!\sigma d \phi^2 \big)\,,
\ea
where
\be
f= 1  - \frac{2mr}{\rho^2}+ \frac{r^2}{l^2}\,,\quad \rho^2=r^2+l^2\,.
\ee
Note that whereas the black brane limits discussed in the previous appendix yield asymptotically flat metrics, this limit retains the asymptotically AdS structure of the spacetime.

Let us next consider the doubly spinning black hole of minimal gauged supergravity studied in Sec.~\ref{SecSUGRA}. For concreteness and future reference we shall perform the
hyperboloid membrane limit in the $\psi$-direction, that is send $b\to l$, in a coordinate system $(t,r,\theta,\phi_R, \psi)$ where the coordinate $\phi_R$ rotates at infinity.
Starting from the metric \eqref{SUGRAmetric} with $\nu,\omega, d\gamma^2$ given by \eqref{nuomega2}, \eqref{gamma4}, we apply the following
substitution analogous to \eqref{eqn:hyperboliclimit}:
\begin{align}
	\cos\theta
	&=
	\sqrt{\Xi_b} \cosh(\sigma/2)\,.
	\label{eqn:}
\end{align}
Note that while \eqref{eqn:hyperboliclimit} ``zooms in" on $\theta = 0$, this substitution zooms in on $\theta=\pi/2$, so that we may take $b \to l$ instead of $a\to l$.

Upon the limit $b\to l$ we then have $\nu \to \nu_m$, $\omega\to \omega_m$, $d\gamma^2\to d\gamma^2_m$, where
\ba
\nu_m&=&\frac{a}{l}dt+ld\phi_R\,,\nn\\
\omega_m&=&dt-\frac{a}{\Xi_a}d\phi_R-l\cosh^2(\sigma/2) d\psi\,,\nn\\
d\gamma^2_m&=&\frac{1}{\Xi_a}\Bigl[(r^2+a^2)d\phi^2_R+\frac{2a}{l^2}(r^2+a^2)dtd\phi_R\nonumber\\
&&\ \ \ \ -\frac{dt^2}{l^2}\bigl(\rho^2-(r^2+a^2)\frac{a^2}{l^2}\bigr)\Bigr]\nonumber\\
&&+\cosh^2(\sigma/2)\rho^2\Bigl[d\psi^2-\frac{dt^2}{l^2}\Bigr]\,.
\ea
The hyperboloid membrane solution then reads
\ba\label{SUGRAmetricMebrane}
ds^2&=& d\gamma^2_m - \frac{2q \nu_m \omega_m}{\rho^2}+\frac{f_m \omega^2_m}{\rho^4} + \frac{\rho^2 dr^2}{\Delta_m}\qquad\ \nn\\
&&+ \frac{\rho^2\sinh^2\!(\sigma/2)}{\Xi_a\cosh^2(\sigma/2)+1}\frac{d\sigma^2}{4}\,,\quad
\nn\\
A &=& \frac{\sqrt{3}q\omega_m}{\rho^2}\,,
\ea
where
\ba
\Delta_m &=& \frac{(r^2+a^2)\rho^4/l^2+q^2+2alq}{r^2}-2m\,,\quad \rho^2=r^2+l^2\,,
\nn\\
f &=& 2m\rho^2 -q^2+\frac{2aq}{l}\rho^2\,,\quad
\Xi_a = 1-\frac{a^2}{l^2}\,.
\ea
This is a consistent solution of the Einstein--Maxwell-AdS equations in 5 dimensions.

Obviously, no additional hyperboloid membrane limit can be taken in the $a$-direction [neither it were  possible to take simultaneously two hyperboloid membrane limits of the original equal spinning metric]. However, the form of the metric \eqref{SUGRAmetricMebrane} suggests that an additional super-entropic limit can be taken. Indeed,
beginning with this metric and rescaling $\phi_R = \Xi_a \varphi$ in the usual way, followed by the $a\to l$ limit, we have $\nu_m\to dt$,
\ba
\omega_m\to \omega_s&=&dt-ld\varphi -l\cosh^2(\sigma/2) d\psi\,,\nn\\
d\gamma^2_m\to d\gamma_s&=&
\frac{1}{l^2}\Bigl(2l\rho^2dtd\varphi-(\rho^2+l^2)dt^2\Bigr)\nn\\
&&+\rho^2\cosh^2(\sigma/2)\Big(d\psi^2-\frac{dt^2}{l^2}\Bigr)\,.
\ea
Hence we recovered the following super-entropic hyperboloid membrane solution of minimal gauged gravity:
\ba\label{SUGRAmetricMebrane2}
ds^2&=& d\gamma^2_s - \frac{2q dt \omega_s}{\rho^2}\!+\!\frac{f_s \omega^2_s}{\rho^4} \!+\! \frac{\rho^2 dr^2}{\Delta_s}
\!+\! \frac{\rho^2\sinh^2\!(\sigma/2)}{4}d\sigma^2\,,\nn\\
A &=& \frac{\sqrt{3}q\omega_m}{\rho^2}\,,
\ea
where
\be
\Delta_s = \frac{\rho^6/l^2+q^2+2l^2q}{r^2}-2m\,,\ f_s = 2m\rho^2 -q^2+2q\rho^2\,
\ee
demonstrating that the super-entropic and hyperboloid membrane limits are compatible. Furthermore, they are commutative: one is free to take the limits in either order and the resultant solution will yield the metric \eqref{SUGRAmetricMebrane2}.  We leave a further analysis of the properties of this solution 
for future work.

\section{Uniqueness of the Rotating Frame}
\label{AppC}

Throughout our analysis we have employed rotating-at-infinity coordinates when taking the super-entropic ultraspinning limit. Here we will examine the uniqueness of the choice of rotating frame, discussing for simplicity the four-dimensional Kerr-AdS case.

Let us begin with the Kerr-AdS solution written in the standard Boyer--Lindquist form, given by \eqref{KNADS2}, \eqref{KerrSigmaa} in the main text.
In this form, the metric is already written in `rotating coordinates', characterized by $\Omega_\infty = -a/l^2$.  The fact that these coordinates are `rotating' is crucial for the super-entropic limit---working in non-rotating coordinates leads to a singular limit.  We can ask, though, what restrictions (if any) are there on the rotating coordinates we use?  That is, are there other frames (besides that characterized by $\Omega_\infty = -a/l^2$) in which it is possible to perform the super-entropic limit?  Let us begin to answer this question by writing the metric in `non-rotating coordinates' by transforming,
\be
\Phi = \phi + \frac{a}{l^2}t\,,
\ee
where $\Phi$ is the non-rotating coordinate.  We find,
\ba\label{KNADS3}
ds^2&=&-\frac{\Delta_a}{\Sigma_a}\left[\left(1+\frac{a^2\sin^2\!\theta}{l^2 \Xi} \right)dt-\frac{a\sin^2\!\theta}{\Xi}d\Phi\right]^2
\nn\\
&+&\frac{\Sigma_a}{\Delta_a} dr^2+\frac{\Sigma_a}{S}d\theta^2\nonumber\\
&+&\frac{S\sin^2\!\theta}{\Sigma_a}\left[\left(a+ \frac{a}{l^2}\frac{r^2+a^2}{\Xi} \right)  dt-\frac{r^2+a^2}{\Xi}d\Phi\right]^2\nn\,.
\ea
It is now be clear that the limit cannot be directly taken in the non-rotating coordinates: the $g_{tt}$ and $g_{t\Phi}$ components of the metric are singular in the $a\to l$ limit and cannot be made finite through our rescaling of $\phi$.  There appears to be two possible ways to fix this: one could simply re-scale $t$ as $t \to \Xi t$ while simultaneously taking $\phi \to \Xi \phi$  or we could transform to a rotating frame and then take $\phi \to \Xi \phi$.   It turns out that the first method does not work (it leads to a singular metric) and so transforming to a rotating frame is essential.

Now, starting from the non-rotating metric let us transform to an {\it arbitrary} rotating frame via the transformation,
\be
\varphi = \Phi - x\frac{a}{l^2}t\,,
\ee
where $x$ is (for now) an arbitrary parameter.  Note that with the choice $x=1$ eq. \eqref{KNADS2} is recovered.  We then have for the metric in rotating-at-infinity coordinates,
\ba\label{KNADS4}
ds^2&=&-\frac{\Delta_a}{\Sigma_a}\left[\left(1+\frac{a^2\sin^2\!\theta}{l^2 \Xi}(1-x) \right) dt-\frac{a\sin^2\!\theta}{\Xi}d\varphi\right]^2
\nn\\
&+&\frac{S\sin^2\!\theta}{\Sigma_a}\left[\left(1+ \frac{r^2+a^2}{l^2\Xi}(1-x) \right) a dt-\frac{r^2+a^2}{\Xi}d\varphi\right]^2\nonumber\\
&+&\frac{\Sigma_a}{\Delta_a} dr^2+\frac{\Sigma_a}{S}d\theta^2\,.
\ea
Considering this metric we see that $g_{tt}$ and $g_{t\varphi}$ components can be made finite with the choice
\be
x=1+ y\Xi+o(\Xi)\,,
\ee
where $y$ is a parameter, with  $y=0$ yielding the coordinates we have used throughout the paper, and $o(\Xi)$ denotes terms of higher order in $\Xi$.
Note that we cannot have $y \propto \Xi^{-1}$ or the argument will not work.  We then have, in these coordinates,
\be \Omega_\infty  = -\frac{a}{l^2}\left( 1 + \Xi y \right)\,. \ee
This result tells us that we do face some restrictions in our choice of coordinates.  For example, it is not possible to perform the super-entropic limit if one begins in coordinates that rotate at infinity with $\Omega_\infty = -2a/l^2$ since this would require $y = 1/\Xi$, which is not valid.  Now we must ask: when we perform the super-entropic limit in coordinates with an arbitrary (but valid) choice of $y$, how is the result related to our standard choice of $y=0$?

The answer is that different values of $y$ correspond simply to coordinate transformations of the solutions discussed in this paper -- there is nothing qualitatively different about the solution.
To see this consider the transformation we made to the rotating frame
\be
\varphi = \Phi - x\frac{a}{l^2}t=\Phi - \frac{a}{l^2}t - y\Xi \frac{a}{l^2}t\,.
\ee
Now recall that, at this point, when taking the super-entropic limit, we rescale $\varphi$ via $\varphi = \Xi \psi$ and then take $a\to l$.  So, with a non-vanishing $y$ we have:
\be
\psi = \frac{\varphi}{\Xi} = \frac{\Phi -\frac{a}{l^2}t}{\Xi} -  y \frac{a}{l^2}t\ \
 \mathrel{\stackrel{\makebox[0pt]{\mbox{\normalfont\small \text{$a \!\to\! l$}}}}{=}} \,\,\, \psi_{\text{SE}} -\frac{y}{l}t\,,
\ee
where $\psi_{\text{SE}}$ denotes the azimuthal coordinate from the super-entropic solutions.  So beginning in other rotating-coordinate systems just turn out to yield a simple coordinate transformation applied to the solution we have already obtained.

 We need to move to a rotating coordinate system because otherwise we will have a divergence in $g_{tt}$ and $g_{t\phi}$.  While there is some freedom in the choice of starting frame, we cannot perform the super-entropic limit from any rotating frame whatsoever.  When an appropriate coordinate system is chosen, however, we always recover the `standard' super-entropic solution (up to a simple coordinate transformation).


\providecommand{\href}[2]{#2}\begingroup\raggedright\endgroup

\end{document}